\documentclass{article}
\usepackage{amssymb}
\usepackage{hyperref}
\usepackage[toc,page]{appendix}

\hypersetup{linktocpage}
\newtheorem{theorem}{Theorem}
\newtheorem{acknowledgement}[theorem]{Acknowledgement}

\newtheorem{example}[theorem]{Example}

\newenvironment{proof}[1][Proof]{\noindent\textbf{#1.} }{\ \rule{0.5em}{0.5em}}
\input{tcilatex}
\begin{document}

\title{Nonequilibrium superfield and lattice Weyl transform transport
approach to quantum Hall effect}
\author{Felix A. Buot \\
Laboratory of Computational Functional Materials, \\
Nanoscience and Nanotechnology (LCFMNN), \\
Department of Physics, University of San Carlos\\
Talamban, Cebu City 6000, Philippines \\
Center for Theoretical Condensed Matter Physics (CTCMP), \\
Cebu Normal University, Cebu City 6000, Philippines \\
C\&LB Research Institute, Carmen, Cebu 6005, Philippines}
\maketitle

\begin{abstract}
We derive the topological Chern number of the integer quantum Hall effect in
electrical conductivity, using Buot's superfield and lattice Weyl transform
nonequilibirum quantum transport formalism. The method is naturally
straightforward, appropriate for treating nonequilibirum systems acted on by
external electromagnetic fields. We have identified the topological
invariant in the effective $\left( \vec{p},\vec{q};E,t\right) $-phase space
via the nonequilibrium quantum transport equation, generally not to
first-oder in electric field but to first-order in the gradient expansion.
We have also derive the Kubo current-current correlation for the Hall
current as a by-product of our new transport approach. The Berry curvature
related to orbital magnetic moment is also calculated.

\pagebreak
\end{abstract}

\tableofcontents

\section{Introduction}

Here, we consider the Hall conductance in a two-dimensional periodic
potential, a nonequilibrium system driven by an electromotive force. The
quantization of Hall conductance in a two-dimensional periodic potential was
first explained by Thouless, Kohmoto, Nightingale, and den Nijs (TKNN) \cite%
{tknn} using the Kubo current-current correlation. Similar approach was
employed by Streda \cite{streda}. Earlier, Laughlin \cite{laughlin}, and
later Halperin \cite{halperin}, study the effects produced by changes in the
vector potential on the states at the edges of a finite system, where
quantization of the conductance is made explicit, but it was not obvious
that the result is insensitive to boundary conditions. In contrast, the use
of Kubo formula by TKNN is for bulk two-dimensional conductors.

These theoretical studies were motivated by the Nobel Prize winning
experimental discovery of von Klitzing, Dorda, and Pepper \cite{kdp} on the
quantization of the Hall conductance of a two-dimensional electron gas in a
strong magnetic field. The strong magnetic field basically provides the
gapped energy structure for the experiments. In the TKNN approach, periodic
potential in crystalline solid is being treated. A strong magnetic field is
not needed to provide the gapped energy structure in their theory, only
peculiar, gapped energy-band structures. In principle, in the presence of
electric field the discrete Landau levels is replaced by the \textit{unstable%
} discrete Stark ladder-energy levels \cite{zener, wannier}.

In this paper, we simply make use of the gapped energy-band structure of
solids under external electric field. We employ the lattice Weyl
transformation technique to go from the bare Hamiltonian to effective or
renormalized crystal Hamiltonian, which also provides a rigorous
justification of the usual \textit{ansatz} of substituting the coordinate
operator $\hat{r}$ by $i\frac{\partial }{\partial \vec{p}}$ in $\vec{p}$%
-space (crystal momentum), as well as the crystal group velocity $\vec{v}%
_{g}=\frac{\partial H}{\partial \vec{p}}$. We then employ the real-time
superfield and lattice Weyl transform nonequilibrium Green's function
(SFLWT-NEGF) \cite{buotbook} quantum transport formalism of Buot \cite{trHn,
bj} in the first-order gradient expansion to derive the topological Chern
number of the integer quantum Hall effect (IQHE) for two-dimensional
systems, as an integral multiple of quantum conductance, also known as
contact conductance in mesoscopic physics \cite{buotbook}.

We find that the quantization of Hall effect occurs strictly not to first
order in the electric field \textit{per se} but rather to first-order
gradient expansion in the quantum transport equation. The Berry connection
and Berry curvature is the fundamental physics \cite{fego} behind the
demonstration of the exact quantization of Hall conductance in units of $%
\frac{e^{2}}{h}$, which also happens to coincide with the source and drain 
\textit{contact} conductance per spin in a closed circuit of mesoscopic
quantum transport \cite{buotbook}.

The new method employed in this paper employs neither the conventional use
of Kubo formula originally employed by TKNN \cite{tknn} nor the use of
retarded Green's function in linear response theory of equilibrium systems.
We have identified the topological invariant in $\left( p.q;E.t\right) $%
-space quantum transport to be given by%
\[
\frac{1}{\left( 2\pi \hbar \right) }\int \int \int d\mathcal{\vec{K}}_{x}d%
\mathcal{\vec{K}}_{y}dt\ \left[ \frac{\partial ^{\left( a\right) }}{\partial 
\mathcal{\vec{K}}_{x}}\frac{\partial ^{\left( b\right) }}{\partial \mathcal{%
\vec{K}}_{y}}-\frac{\partial ^{\left( a\right) }}{\partial \mathcal{\vec{K}}%
_{y}}\frac{\partial ^{\left( b\right) }}{\partial \mathcal{\vec{K}}_{x}}%
\right] H^{\left( a\right) }\left( \mathcal{\vec{K}},\mathcal{E}\right)
\left( -iG^{<\left( b\right) }\left( \mathcal{\vec{K}},\mathcal{E}\right)
\right) \text{.} 
\]%
Here, the uniform electric field, $\vec{F}$, is in the $x$-direction and 
\[
\mathcal{\vec{K}}=\vec{p}+e\vec{F}t. 
\]%
Moreover, the formula is also applicable to gapped Landau-level structure of
a free electron gas in intense magnetic field \cite{kdp} since the variable $%
\mathcal{\vec{K}}$ can incorporates the external vector potential if
present. Note that the change of variables from $\mathcal{\vec{K}}$ to $\vec{%
p}$ in the integration over the whole Brillouin zone has a Jacobian unity.

The above formula in $\mathcal{\vec{K}}$-space is given only for one
occupied energy band. Summation over all separated (gapped) energy bands can
be accommodated in the above formula by adding the pertinent summation
symbol over the occupied bands, with $H^{\left( a\right) }$ belonging to
respective bands.

To the author's knowledge, this is the first time the real-time-dependent
quantum superfield nonequilibrium transport is employed to derive
topological invariant in phase space of Chern condensed matter system.
Clearly, a nonequilibrium quantum transport approach is called for since any
system under a external electric field is a nonequilibrium system where
current flows. In our unique approach, we have derived the Kubo
current-current formula strictly from real-time nonequilibrium quantum
superfield theory of transport physics adapted to time-dependent electric
fields. In contrast, the Kubo current-current approach approximates a
nonequilibrium situation as a disturbed equilibrium system under an electric
field. In this paper, the orbital magnetic moment, and its related Berry
curvature are also calculated.

\section{The Wigner distribution function and density matrix}

In the nonequilibrium many-body Green's function technique, the principal
quantities of interest are the \textquotedblleft reduced\textquotedblright\
or single-particle correlation functions defined as

\begin{equation}
-iG^{<}\left( 1,2\right) =Tr\left[ \rho _{H}\left( \psi _{H}^{\dagger
}\left( 2\right) \right) \psi _{H}\left( 1\right) \right] =\left\langle
\left( \psi _{H}^{\dagger }\left( 2\right) \right) \psi _{H}\left( 1\right)
\right\rangle \text{,}  \label{2pointCorr}
\end{equation}%
where $\psi _{H}\left( 1\right) $ and $\psi _{H}^{\dagger }\left( 2\right) $
are the particle annihilation and creation operators in the Heisenberg
representation, respectively. The indices $1$ and $2$ subsume all space-time
indices and other quantum-label indices. As we shall see in what follows,
the equation for $G^{<}\left( 1,2\right) $ plays a principal role in
particle quantum transport since it stands for particle distribution in
phase space.

If we write the second quantization operator, $\hat{f}$, for the
one-particle $\left( \vec{p},\vec{q};E,t\right) $-phase space distribution
function\footnote{%
The Buot's formulation of discrete lattice Weyl transform is generally based
on a finite field represented by a finite prime modulus number, in all
periodic directions of lattice points on a `torus' obeying modular
arithmetic, closed under addition, multiplication and division, where
division by all nonzero numbers is possible if and only if the modulus is
prime. In the terminology of abstract algebra, the ability to perform
division means that modular arithmetic modulo a prime number forms a finite
field. This specific aspect of Buot's theory has been given a firm
theoretical foundation, discussed in detail in the Concluding section.} as%
\begin{equation}
\hat{f}_{\lambda \lambda ^{\prime }\sigma \sigma ^{\prime }}\left( \vec{p},%
\vec{q};E,t\right) =\dsum\limits_{\vec{v}}e^{\frac{i}{\hbar }\left( 2\vec{p}%
\cdot \vec{v}+E\tau \right) }\psi _{\lambda \sigma }^{\dagger }\left( \vec{q}%
+\vec{v},t+\frac{\tau }{2}\right) \psi _{\lambda ^{\prime }\sigma ^{\prime
}}\left( \vec{q}-\vec{v},t-\frac{\tau }{2}\right) \text{,}  \label{distop}
\end{equation}%
where $\lambda $ label the band index and $\sigma $ the spin index [here we
drop the Heisenberg representation subscripts $H$ for economy of indices],
then upon taking the average%
\begin{equation}
\left\langle \hat{f}_{\lambda \lambda ^{\prime }\sigma \sigma ^{\prime
}}\left( \vec{p},\vec{q};E,t\right) \right\rangle =\sum\limits_{\vec{v}}e^{%
\frac{i}{\hbar }\left( 2\vec{p}\cdot \vec{v}+E\tau \right) }\left\langle
\psi _{\lambda \sigma }^{\dagger }\left( \vec{q}+\vec{v},t+\frac{\tau }{2}%
\right) \psi _{\lambda ^{\prime }\sigma ^{\prime }}\left( \vec{q}-\vec{v},t-%
\frac{\tau }{2}\right) \right\rangle \text{.}  \label{lwtrho}
\end{equation}%
Upon employing a four-dimensional notation: $p=\left( \vec{p},E\right) $ and 
$q=\left( \vec{q},t\right) $, we obtain particle distribution function $\rho
_{\lambda \lambda ^{\prime }\sigma \sigma ^{\prime }}\left( p,q\right) ,$%
\begin{equation}
\rho _{\lambda \lambda ^{\prime }\sigma \sigma ^{\prime }}\left( p,q\right)
=\left\langle \hat{f}_{\lambda \lambda ^{\prime }\sigma \sigma ^{\prime
}}\left( p,q\right) \right\rangle \text{.}  \label{average}
\end{equation}%
Equation (\ref{lwtrho}) is indeed the lattice Weyl transform of the density
matrix operator $\hat{\rho}$ as%
\begin{eqnarray}
\rho _{\lambda \lambda ^{\prime }\sigma \sigma ^{\prime }}\left( p,q\right)
&=&\left\langle \hat{f}_{\lambda \lambda ^{\prime }\sigma \sigma ^{\prime
}}\left( p,q\right) \right\rangle  \nonumber \\
&=&\sum\limits_{v}\exp \left( \frac{2i}{\hbar }p\cdot v\right) \left\langle
q-v;\lambda ^{\prime },\sigma ^{\prime }\right\vert \hat{\rho}\left\vert
q+v;\lambda ,\sigma \right\rangle \text{,}  \label{lwtdop}
\end{eqnarray}%
where the RHS is the lattice Weyl transform (LWT) of the density matrix
operator, which is identical to the LWT of $-iG^{<}\left( 1,2\right) $. For
convenience in our $4$-dimensional notation, we have also change variable of
continuous time $t$ from $\frac{t}{2}$ to $t$ resulting in an overall factor 
$2$ in the exponential, so that $v=\left( \vec{v},\tau \right) $ in the
discrete summation over $\vec{v}$ and continous integration over $\tau $.
For convenience, we just used the summation symbol in Eq. (\ref{lwtdop}) to
mean summation and integration over the respective discrete and continuous
variables.

Thus, the expectation value of one-particle operator $\hat{A}$ can be
calculated \textit{in phase-space} like the classical averages using a
distribution function,%
\begin{equation}
Tr\left( \hat{\rho}\ \hat{A}\right) :=\left\langle \hat{A}\right\rangle
=\sum\limits_{p,q;\lambda \lambda ^{\prime }\sigma \sigma ^{\prime
}}A_{\lambda \lambda ^{\prime }\sigma \sigma ^{\prime }}\left( p,q\right)
\rho _{\lambda ^{\prime }\lambda \sigma ^{\prime }\sigma }\left( p,q\right) 
\text{,}  \label{productTr}
\end{equation}%
clearly exhibiting the trace of binary operator product as a trace of the
product of their respective LWT's. This general observation is \textit{%
crucial} in most of the calculations that follows.

The Wigner distribution function $f_{W}\left( \vec{p},\vec{q},t\right) $
maybe given by%
\begin{equation}
f_{W}\left( \vec{p},\vec{q},t\right) =\frac{1}{2\pi \hbar }\int dE\left(
-iG^{<}\left( \vec{p},\vec{q};E,t\right) \right) \text{.}
\label{wignerGless}
\end{equation}

We further note that, 
\begin{eqnarray}
\hat{\rho}\left( t\right) &=&e^{-\frac{i}{\hbar }\hat{H}t}\hat{\rho}\left(
0\right) e^{\frac{i}{\hbar }\hat{H}t},  \nonumber \\
&=&\hat{U}\left( t\right) \hat{\rho}\left( 0\right) \hat{U}^{\dagger }\left(
t\right) \text{,}  \label{timedepend}
\end{eqnarray}%
provides the major time dependence in the transport equation that follows.

In Appendix $A$, we give the renormalized dynamical variables appropriate
for solid-state problems.

\subsection{Phase factors in space and time translation operators}

In Appendix $B$, we give a detailed derivation the \textit{phase factors}
entering in the time and space translation operators via the action
principle, namely, 
\begin{equation}
\hat{T}\left( \vec{q}\right) \hat{T}\left( t\right) =\exp \frac{i}{\hbar }%
\left( \hat{P}\cdot \vec{q}-\hat{\xi}t\right) ,  \label{Tspcetme}
\end{equation}%
where $\hat{P}$ is the momentum operator, $-i\hbar \nabla _{\vec{q}}\psi =%
\hat{P}\ \psi $, and $\hat{\xi}$ is the energy variable operator, explicitly
given by $\hat{\xi}=i\hbar \frac{\partial }{\partial t}$, since in the Schr%
\"{o}dinger equation, $i\hbar \frac{\partial }{\partial t}\psi =\hat{H}\
\psi $. Equation (\ref{Tspcetme}) is showing $\hat{T}\left( t\right) $ and $%
\hat{T}\left( \vec{q}\right) $, respectively, as a product on the left side
since by virtue of the Baker--Campbell--Hausdorff formula, $\hat{T}\left( 
\vec{q}\right) $ and $\hat{T}\left( t\right) $ commute and therefore equal
to the right side of the equation. We obtain the following relations,%
\begin{equation}
\frac{\partial \hat{T}\left( -\vec{q}\right) }{\partial t}=\frac{i}{\hbar }%
\left[ \mathcal{H},\hat{T}\left( -\vec{q}\right) \right] =\frac{i}{\hbar }e%
\vec{F}\cdot \left( -\vec{q}\right) \ \hat{T}\left( -\vec{q}\right) \text{,}
\label{eq8}
\end{equation}%
where $\vec{F}$ is the electric field and $e$ is the electric charge.
Similarly, we have%
\begin{equation}
\frac{\partial \hat{T}\left( t\right) }{\partial \vec{q}}=\frac{i}{\hbar }%
\left[ \mathcal{\vec{K}},\hat{T}\left( t\right) \right] =\left( -\frac{i}{%
\hbar }e\vec{F}t\right) \ \hat{T}\left( t\right) .  \label{eq3}
\end{equation}%
Equations (\ref{eq8}) and (\ref{eq3}) suggest that in the presence of
electric field, gauge invariant quantities that are \textit{displaced} in
space and time acquires a generalized Peierls phase factors \cite{bj}. As an
example, for nonequilibrium translational symmetric and steady-state
condition, 
\begin{equation}
\left\langle \vec{q}_{1},t_{1}\right\vert \hat{H}^{\left( 1\right)
}\left\vert \vec{q}_{2},t_{2}\right\rangle \Longrightarrow e^{-i\frac{e}{%
\hbar }\vec{F}t\cdot \left( \vec{q}_{1}-\vec{q}_{2}\right) }e^{-i\frac{e}{%
\hbar }\vec{F}\cdot \vec{q}\left( t_{1}-t_{2}\right) }H^{\left( 1\right)
}\left( \vec{q}_{1}-\vec{q}_{2},t_{1}-t_{2}\right) \text{,}
\label{peierlsPhase}
\end{equation}%
where%
\begin{eqnarray*}
\vec{q} &=&\frac{1}{2}\left( \vec{q}_{1}+\vec{q}_{2}\right) \text{,} \\
t &=&\frac{1}{2}\left( t_{1}+t_{2}\right) \text{.}
\end{eqnarray*}

Equation (\ref{peierlsPhase}) is a generalization of the well-known Peierls
phase factor in solid-state physics. It is critical in getting the correct
lattice Weyl transform of similar quantities, which are nonlocal functions
of space and time, to phase space. Using the four dimensional notation: $%
p=\left( \vec{p},E\right) $ and $q=\left( \vec{q},t\right) $, the Weyl
transform $A\left( p,q\right) $ of any operator $\mathbf{\hat{A}}$ is
defined by%
\begin{eqnarray}
A_{\lambda \lambda ^{\prime }}\left( p.q\right) &=&\sum\limits_{v}e^{\left( 
\frac{2i}{\hbar }\right) p\cdot v}\left\langle q-v,\lambda \right\vert 
\mathbf{\hat{A}}\left\vert q+v,\lambda ^{\prime }\right\rangle ,  \label{LWT}
\\
&=&\sum\limits_{u}e^{\left( \frac{2i}{\hbar }\right) q\cdot u}\left\langle
p+u,\lambda \right\vert \mathbf{\hat{A}}\left\vert p-u,\lambda ^{\prime
}\right\rangle \text{,}  \label{LWT1}
\end{eqnarray}%
where $\lambda $ and $\lambda ^{\prime }$ stands for other discrete quantum
numbers, e.g., energy band and spin indices. Viewed as a transformation of a
matrix, we see that the Weyl transform of the matrix $\left\langle q^{\prime
},\lambda ^{\prime }\right\vert \mathbf{\hat{A}}\left\vert q^{\prime \prime
},\lambda ^{^{\prime \prime }}\right\rangle $ is given by Eq. (\ref{LWT})
and the lattice Weyl transform of $\left\langle p^{\prime },\lambda ^{\prime
}\right\vert \mathbf{\hat{A}}\left\vert p^{\prime \prime },\lambda ^{\prime
\prime }\right\rangle $ is given by Eq. (\ref{LWT1}). Denoting the operation
of taking the lattice Weyl transform by the symbol $\mathcal{W}$ then it is
easy to see that the lattice Weyl transform of total partial derivatives, 
\begin{eqnarray}
\mathcal{W}\left( \frac{\partial }{\partial q^{\prime }}+\frac{\partial }{%
\partial q^{\prime \prime }}\right) \left\langle q^{\prime },\lambda
^{\prime }\right\vert \mathbf{\hat{A}}\left\vert q^{\prime \prime },\lambda
^{^{\prime \prime }}\right\rangle &=&\frac{\partial }{\partial q}%
\sum\limits_{v}e^{\left( \frac{2i}{\hbar }\right) p\cdot v}\left\langle
q-v,\lambda ^{\prime }\right\vert \mathbf{\hat{A}}\left\vert q+v,\lambda
^{\prime \prime }\right\rangle ,  \nonumber \\
&=&\frac{\partial }{\partial q}A_{\lambda ^{\prime }\lambda ^{\prime \prime
}}\left( p.q\right) \text{.}  \label{sumq}
\end{eqnarray}%
Similarly 
\begin{eqnarray}
\mathcal{W}\left( \frac{\partial }{\partial p^{\prime }}+\frac{\partial }{%
\partial p^{\prime \prime }}\right) \left\langle p^{\prime },\lambda
^{\prime }\right\vert \mathbf{\hat{A}}\left\vert p^{\prime \prime },\lambda
^{^{\prime \prime }}\right\rangle &=&\frac{\partial }{\partial p}%
\sum\limits_{u}e^{\left( \frac{2i}{\hbar }\right) q\cdot u}\left\langle
p+u,\lambda \right\vert \mathbf{\hat{A}}\left\vert p-u,\lambda ^{\prime
}\right\rangle ,  \nonumber \\
&=&\frac{\partial }{\partial p}A_{\lambda ^{\prime }\lambda ^{\prime \prime
}}\left( p.q\right) \text{.}  \label{sump}
\end{eqnarray}%
Note that the derivatives on the LHS of Eqs. (\ref{sumq}) and (\ref{sump})
obviously operate only on the wavefunctions or state vectors not on the
operator.

Writing Eq. (\ref{LWT}) explicitly, we have%
\begin{equation}
A_{\lambda \lambda ^{\prime }}\left( \vec{p}.\vec{q};E,t\right)
=\sum\limits_{\vec{v};\tau }e^{\left( \frac{2i}{\hbar }\right) \vec{p}\cdot 
\vec{v}}e^{\left( \frac{i}{\hbar }\right) E\tau }\left\langle \vec{q}-\vec{v}%
;t-\frac{\tau }{2},\lambda \right\vert \mathbf{\hat{A}}\left\vert \vec{q}+%
\vec{v};t+\frac{\tau }{2},\lambda ^{\prime }\right\rangle \text{.}
\label{LWT2}
\end{equation}%
Using the form of matrix elements in Eq. (\ref{peierlsPhase}), we have%
\begin{eqnarray}
&&\left\langle \vec{q}-\vec{v};t-\frac{\tau }{2},\lambda \right\vert \mathbf{%
\hat{A}}\left\vert \vec{q}+\vec{v};t+\frac{\tau }{2},\lambda ^{\prime
}\right\rangle  \nonumber \\
&=&e^{-i\frac{e}{\hbar }\vec{F}t\cdot \left( \vec{q}_{1}-\vec{q}_{2}\right)
}e^{-i\frac{e}{\hbar }\vec{F}\cdot \vec{q}\left( t_{1}-t_{2}\right) }A\left( 
\vec{q}_{1}-\vec{q}_{2},t_{1}-t_{2}\right)  \nonumber \\
&=&e^{i\frac{e}{\hbar }\vec{F}t\cdot \left( 2\vec{v}\right) }e^{i\frac{e}{%
\hbar }\vec{F}\cdot \vec{q}\tau }A_{\lambda \lambda ^{\prime }}\left( \vec{q}%
_{1}-\vec{q}_{2},t_{1}-t_{2}\right) \text{.}  \label{matrix}
\end{eqnarray}%
Thus%
\begin{eqnarray}
A_{\lambda \lambda ^{\prime }}\left( \vec{p}.\vec{q};E,t\right)
&=&\sum\limits_{\vec{v};\tau }e^{\left( \frac{2i}{\hbar }\right) \vec{p}%
\cdot \vec{v}}e^{\left( \frac{i}{\hbar }\right) E\tau }e^{i\frac{e}{\hbar }%
\vec{F}t\cdot \left( 2\vec{v}\right) }e^{i\frac{e}{\hbar }\vec{F}\cdot \vec{q%
}\tau }A_{\lambda \lambda ^{\prime }}\left( \vec{q}_{1}-\vec{q}%
_{2},t_{1}-t_{2}\right) ,  \nonumber \\
&=&\sum\limits_{\vec{v};\tau }e^{\left( \frac{2i}{\hbar }\right) \left( \vec{%
p}+e\vec{F}t\right) \cdot \vec{v}}e^{\left( \frac{i}{\hbar }\right) \left(
E+e\vec{F}\cdot \vec{q}\right) \tau }A_{\lambda \lambda ^{\prime }}\left( 
\vec{q}_{1}-\vec{q}_{2},t_{1}-t_{2}\right) ,  \nonumber \\
&=&A_{\lambda \lambda ^{\prime }}\left( \left( \vec{p}+e\vec{F}t\right)
;\left( E+e\vec{F}\cdot q\right) \right) ,  \nonumber \\
&=&A_{\lambda \lambda ^{\prime }}\left( \mathcal{\vec{K}};\mathcal{E}\right) 
\text{.}  \label{matrixphase}
\end{eqnarray}%
Hence the expected dynamical variables in the phase space including the time
variable occurs in particular combinations of $\mathcal{\vec{K}}$ and $%
\mathcal{E}$. Therefore, besides \ the crystal momentum varying in time as 
\begin{equation}
\mathcal{\vec{K}}=\vec{p}+e\vec{F}t\text{,}  \label{shiftedK}
\end{equation}%
the energy variable vary with $\vec{q}$ as 
\begin{equation}
\mathcal{E}=E+e\vec{F}\cdot \vec{q}.  \label{shiftedE}
\end{equation}%
In effect we have unified the use of \textit{scalar potential} and \textit{%
vector potential} for a system under uniform electric fields. Thus,%
\begin{eqnarray}
\frac{\partial \mathcal{\vec{K}}}{\partial t} &=&e\vec{F},\text{ \ \ }\frac{%
\partial }{\partial t}=\frac{\partial \mathcal{\vec{K}}}{\partial t}\cdot 
\frac{\partial }{\partial \mathcal{\vec{K}}}=e\vec{F}\cdot \frac{\partial }{%
\partial \mathcal{\vec{K}}}\text{,}  \label{dt} \\
\frac{\partial \mathcal{E}}{\partial \vec{q}} &=&e\vec{F},\text{ \ \ }\frac{%
\partial }{\partial \vec{q}}=\frac{\partial \mathcal{E}}{\partial \vec{q}}%
\frac{\partial }{\partial \mathcal{E}}=e\vec{F}\frac{\partial }{\partial 
\mathcal{E}}\text{,}  \label{dE} \\
\frac{\partial }{\partial \mathcal{\vec{K}}} &=&\frac{\partial \mathcal{E}}{%
\partial \mathcal{\vec{K}}}\frac{\partial }{\partial \mathcal{E}}=\vec{v}_{g}%
\frac{\partial }{\partial \mathcal{E}}\text{,}  \label{vg}
\end{eqnarray}%
where $v_{g}$ is the group velocity. The LWT of the effective or
renormalized lattice Hamiltonian $\mathcal{H}_{eff}\leftrightarrows H\left( 
\vec{p},\vec{q};E,t\right) $ can therefore be analyzed on $\left( \mathcal{%
\vec{K}},\mathcal{E}\right) $-space as%
\begin{equation}
H\left( \vec{p},\vec{q};E,t\right) \Longrightarrow H\left( \mathcal{\vec{K}},%
\mathcal{E}\right) .  \label{HinkappaE}
\end{equation}%
The last line is by virtue of Eqs. (\ref{shiftedK})- (\ref{shiftedE}). Of
course in the absence of the electric field, the dependence in phase space
becomes the familiar $H\left( \vec{p},\mathcal{\omega }\right) $ for
translationally symmetric and steady-state system. But with $\vec{F}\neq 0$
all gauge invariant quantities are functions of $\left( \mathcal{\vec{K}},%
\mathcal{E}\right) $ such as the \textit{electric Bloch function \cite{bj}}
or Houston wavefunction [\cite{zener}] and \textit{electric Wannier function}%
, i.e., the electric-field dependent generalization of Wannier function. In
particular, the Weyl transform of a commutator,%
\begin{equation}
\mathcal{W}\left[ H,G^{<}\right] =\sin \Lambda \text{,}  \label{weylcommutr}
\end{equation}%
where $\Lambda $ is the Poisson bracket operator. We can therefore write the
Poisson bracket operator $\Lambda $, as%
\begin{eqnarray}
\Lambda &=&\frac{\hbar }{2}\left[ \frac{\partial ^{\left( a\right) }}{%
\partial t}\frac{\partial ^{\left( b\right) }}{\partial \mathcal{E}}-\frac{%
\partial ^{\left( a\right) }}{\partial \mathcal{E}}\frac{\partial ^{\left(
b\right) }}{\partial t}\right] ,\text{ }  \nonumber \\
&=&\frac{\hbar }{2}\frac{\partial \mathcal{\vec{K}}}{\partial t}\cdot \left[ 
\frac{\partial ^{\left( a\right) }}{\partial \mathcal{\vec{K}}}\frac{%
\partial ^{\left( b\right) }}{\partial \mathcal{E}}-\frac{\partial ^{\left(
a\right) }}{\partial \mathcal{E}}\frac{\partial ^{\left( b\right) }}{%
\partial \mathcal{\vec{K}}}\right] ,  \nonumber \\
&=&\frac{\hbar }{2}e\vec{F}\cdot \left[ \frac{\partial ^{\left( a\right) }}{%
\partial \mathcal{\vec{K}}}\frac{\partial ^{\left( b\right) }}{\partial 
\mathcal{E}}-\frac{\partial ^{\left( a\right) }}{\partial \mathcal{E}}\frac{%
\partial ^{\left( b\right) }}{\partial \mathcal{\vec{K}}}\right] \text{, }
\label{poissonBracket}
\end{eqnarray}%
on $\left( \mathcal{\vec{K}},\mathcal{E}\right) $-phase space.

\section{SFLWT-NEGF transport equation}

We now make use of the SFLWT-NEGF quantum transport formalism \cite{buotbook}%
. The nonequibrium quantum superfield transport equation for interacting
Bloch electrons under a uniform electric field has been derived in Sec. $VI$
of Buot and Jensen \cite{bj}. The result to first-order gradient expansion
is given in $Sec.VI$, $Eq.(109)$ of their paper. The derivation given there
is more detailed and comprehensive and will not be repeated here. The
discussions above more or less already captured the basic idea.

To recapitulate, in the absence of Cooper pairings and superconducting
transport mechanism, the SFLWT-NEGF phase-space transport equation reduces
to,%
\begin{eqnarray}
\frac{\partial }{\partial t}G^{<}\left( \vec{p},\vec{q};E,t\right) &=&\frac{2%
}{\hbar }\sin \hat{\Lambda}\left\{ H\left( p,q\right) G^{<}\left( p,q\right)
+\Sigma ^{<}\left( p,q\right) \func{Re}G^{r}\left( p,q\right) \right\} 
\nonumber \\
&&+\frac{1}{\hbar }\cos \hat{\Lambda}\left\{ \Sigma ^{<}\left( p,q\right)
A\left( p,q\right) -\Gamma \left( p,q\right) G^{<}\left( p,q\right) \right\} 
\text{.}  \label{eq105}
\end{eqnarray}%
where in the right side of Eq. (\ref{eq105}) the $4$-dimensional notation of
phase space is employed. If we expand Eq. (\ref{eq105}) to first-order in
the gradient, i.e., $\sin \Lambda \simeq \Lambda ,$ Buot and Jensen's
first-order gradient expansion of phase-space transport equation \cite{bj}
can be written in a compact form as 
\begin{eqnarray}
&&\frac{\partial }{\partial t}G^{<}\left( \mathcal{\vec{K}},\mathcal{E}%
\right)  \nonumber \\
&=&-e\vec{F}\cdot \left\{ \frac{\partial }{\partial \mathcal{E}}E_{\alpha
}\left( \mathcal{\vec{K}},\mathcal{E}\right) +\frac{\partial \func{Re}\Sigma
^{r}\left( \mathcal{\vec{K}},\mathcal{E}\right) }{\partial \mathcal{E}}%
\right\} \frac{\partial }{\partial \mathcal{\vec{K}}}G^{<}\left( \mathcal{%
\vec{K}},\mathcal{E}\right)  \nonumber \\
&&+e\vec{F}\cdot \left\{ \frac{\partial }{\partial \mathcal{\vec{K}}}%
E_{\alpha }\left( \mathcal{\vec{K}},\mathcal{E}\right) +\frac{\partial \func{%
Re}\Sigma ^{r}\left( \mathcal{\vec{K}},\mathcal{E}\right) }{\partial 
\mathcal{\vec{K}}}\right\} \frac{\partial }{\partial \mathcal{E}}G^{<}\left( 
\mathcal{\vec{K}},\mathcal{E}\right)  \nonumber \\
&&-e\vec{F}\cdot \left\{ \frac{\partial \Sigma ^{<}\left( \mathcal{\vec{K}},%
\mathcal{E}\right) }{\partial \mathcal{E}}\frac{\partial \func{Re}%
G^{r}\left( \mathcal{\vec{K}},\mathcal{E}\right) }{\partial \mathcal{\vec{K}}%
}\right\}  \nonumber \\
&&+e\vec{F}\cdot \left\{ \frac{\partial \Sigma ^{<}\left( \mathcal{\vec{K}},%
\mathcal{E}\right) }{\partial \mathcal{\vec{K}}}\frac{\partial \func{Re}%
G^{r}\left( \mathcal{\vec{K}},\mathcal{E}\right) }{\partial \mathcal{E}}%
\right\}  \nonumber \\
&&+\frac{1}{\hbar }\left\{ \Sigma ^{<}\left( \mathcal{\vec{K}},\mathcal{E}%
\right) A\left( \mathcal{\vec{K}},\mathcal{E}\right) -\Gamma \left( \mathcal{%
\vec{K}},\mathcal{E}\right) G^{<}\left( \mathcal{\vec{K}},\mathcal{E}\right)
\right\} \text{,}  \label{eqgrad}
\end{eqnarray}%
where $G^{r}\left( \mathcal{\vec{K}},\mathcal{E}\right) $ is the LWT of the
retarded Green's function, $A\left( \mathcal{\vec{K}},\mathcal{E}\right) $
is the spectral function, describing 'scattering-in', and $\Gamma \left( 
\mathcal{\vec{K}},\mathcal{E}\right) $ is the corresponding 'scattering-out'
rate.

\subsection{Balistic transport and diffusion}

We wiill simplify Eq. (\ref{eqgrad}) by neglecting the self-energies, i.e.,
we limit to non-interacting particles. Then we have the following reduced
quantum transport equation,%
\begin{eqnarray}
\frac{\partial }{\partial t}G^{<}\left( \mathcal{\vec{K}},\mathcal{E}\right)
&=&-e\vec{F}\cdot \frac{\partial }{\partial \mathcal{E}}E_{\alpha }\left( 
\mathcal{\vec{K}},\mathcal{E}\right) \frac{\partial }{\partial \mathcal{\vec{%
K}}}G^{<}\left( p,q\right)  \nonumber \\
&&+e\vec{F}\cdot \frac{\partial }{\partial \mathcal{\vec{K}}}E_{\alpha
}\left( \mathcal{\vec{K}},\mathcal{E}\right) \frac{\partial }{\partial 
\mathcal{E}}G^{<}\left( p,q\right) ,  \label{ballistic}
\end{eqnarray}%
which can be written in terms of the Poisson bracket of Eq. (\ref%
{poissonBracket}) as%
\begin{equation}
\frac{\partial }{\partial t}G^{<}\left( \mathcal{\vec{K}},\mathcal{E}\right)
=\frac{2}{\hbar }\frac{\hbar }{2}e\vec{F}\cdot \left[ \frac{\partial
^{\left( a\right) }}{\partial \mathcal{\vec{K}}}\frac{\partial ^{\left(
b\right) }}{\partial \mathcal{E}}-\frac{\partial ^{\left( a\right) }}{%
\partial \mathcal{E}}\frac{\partial ^{\left( b\right) }}{\partial \mathcal{%
\vec{K}}}\right] H^{\left( a\right) }\left( \mathcal{\vec{K}},\mathcal{E}%
\right) G^{<\left( b\right) }\left( \mathcal{\vec{K}},\mathcal{E}\right) 
\text{.}  \label{noninteract}
\end{equation}%
Therefore, integrating with respect to time,%
\begin{equation}
G^{<}\left( \mathcal{\vec{K}},\mathcal{E}\right) =e\vec{F}\cdot \dint dt\ %
\left[ \frac{\partial ^{\left( a\right) }}{\partial \mathcal{\vec{K}}}\frac{%
\partial ^{\left( b\right) }}{\partial \mathcal{E}}-\frac{\partial ^{\left(
a\right) }}{\partial \mathcal{E}}\frac{\partial ^{\left( b\right) }}{%
\partial \mathcal{\vec{K}}}\right] H^{\left( a\right) }\left( \mathcal{\vec{K%
}},\mathcal{E}\right) G^{<\left( b\right) }\left( \mathcal{\vec{K}},\mathcal{%
E}\right) \text{.}  \label{integrate}
\end{equation}%
With the electric field in the $x$-direction, then we have%
\begin{equation}
G^{<}\left( \mathcal{\vec{K}},\mathcal{E}\right) =e\left\vert \vec{F}%
\right\vert \int dt\ \left[ \frac{\partial ^{\left( a\right) }}{\partial 
\mathcal{\vec{K}}_{x}}\frac{\partial ^{\left( b\right) }}{\partial \mathcal{E%
}}-\frac{\partial ^{\left( a\right) }}{\partial \mathcal{E}}\frac{\partial
^{\left( b\right) }}{\partial \mathcal{\vec{K}}_{x}}\right] H^{\left(
a\right) }\left( \mathcal{\vec{K}},\mathcal{E}\right) G^{<\left( b\right)
}\left( \mathcal{\vec{K}},\mathcal{E}\right) \text{.}  \label{xdirection}
\end{equation}

The Hall current in the $y$-direction is thus given by the following
equation, 
\begin{eqnarray}
&&\frac{a^{2}}{\left( 2\pi \hbar \right) ^{2}}\int \int d\mathcal{\vec{K}}%
_{x}d\mathcal{\vec{K}}_{y}\left( \frac{e}{a^{2}}\frac{\partial \mathcal{E}}{%
\partial \mathcal{\vec{K}}_{y}}\right) \left( -iG^{<}\left( \mathcal{\vec{K}}%
,\mathcal{E}\right) \right)  \nonumber \\
&=&e^{2}\left\vert \vec{F}\right\vert \frac{1}{\left( 2\pi \hbar \right) ^{2}%
}\int \int \int \ d\mathcal{\vec{K}}_{x}d\mathcal{\vec{K}}_{y}\ dt  \nonumber
\\
&&\times \frac{\partial \mathcal{E}}{\partial \mathcal{\vec{K}}_{y}}\left[ 
\frac{\partial ^{\left( a\right) }}{\partial \mathcal{\vec{K}}_{x}}\frac{%
\partial ^{\left( b\right) }}{\partial \mathcal{E}}-\frac{\partial ^{\left(
a\right) }}{\partial \mathcal{E}}\frac{\partial ^{\left( b\right) }}{%
\partial \mathcal{\vec{K}}_{x}}\right] H^{\left( a\right) }\left( \mathcal{%
\vec{K}},\mathcal{E}\right) \left( -iG^{<\left( b\right) }\left( \mathcal{%
\vec{K}},\mathcal{E}\right) \right) ,  \nonumber \\
&=&e^{2}\left\vert \vec{F}\right\vert \frac{1}{\left( 2\pi \hbar \right) ^{2}%
}\int \int \int d\mathcal{\vec{K}}_{x}d\mathcal{\vec{K}}_{y}dt  \nonumber \\
&&\times \ \left[ \frac{\partial ^{\left( a\right) }}{\partial \mathcal{\vec{%
K}}_{x}}\frac{\partial ^{\left( b\right) }}{\partial \mathcal{\vec{K}}_{y}}-%
\frac{\partial ^{\left( a\right) }}{\partial \mathcal{\vec{K}}_{y}}\frac{%
\partial ^{\left( b\right) }}{\partial \mathcal{\vec{K}}_{x}}\right]
H^{\left( a\right) }\left( \mathcal{\vec{K}},\mathcal{E}\right) \left(
-iG^{<\left( b\right) }\left( \mathcal{\vec{K}},\mathcal{E}\right) \right) 
\text{.}  \label{eqKE}
\end{eqnarray}%
If we are only interested in linear response we may consider all the
quantities in the integrand to be of zero-order in the electric field,
although this is not necessary if we allow very weak electric field leading
to time dependence being dominated by the time dependence of the density
matrix, as we shall see in what follows.

\subsection{Quantization of Hall conductance}

From Eq. (\ref{eqKE}), we claim that the quantized Hall conductivity for an
occupied energy band is given by 
\begin{eqnarray}
\sigma _{yx} &=&\left( \frac{e^{2}}{h}\right) \frac{1}{\left( 2\pi \hbar
\right) }\int \int \int d\mathcal{\vec{K}}_{x}d\mathcal{\vec{K}}_{y}dt\  
\nonumber \\
&&\times \left[ \frac{\partial ^{\left( a\right) }}{\partial \mathcal{\vec{K}%
}_{x}}\frac{\partial ^{\left( b\right) }}{\partial \mathcal{\vec{K}}_{y}}-%
\frac{\partial ^{\left( a\right) }}{\partial \mathcal{\vec{K}}_{y}}\frac{%
\partial ^{\left( b\right) }}{\partial \mathcal{\vec{K}}_{x}}\right]
H^{\left( a\right) }\left( \mathcal{\vec{K}},\mathcal{E}\right) \left(
-iG^{<\left( b\right) }\left( \mathcal{\vec{K}},\mathcal{E}\right) \right) 
\text{,}  \label{hallcond}
\end{eqnarray}%
and is quantized in units of $\frac{e^{2}}{h}$, i.e., $\sigma _{yx}=\frac{%
e^{2}}{h}%
\mathbb{Z}
$, where $%
\mathbb{Z}
$ is in the domain of integers or the first Chern numbers. In doing the
integration with respect to time, $t$, we need to examine the implicit
time-dependence of the matrix element of $G^{<}$ in the 'pull back'
representation defined below, which is a process of reverting to matrix
element representation.

\section{Reverting to matrix elements}

Now to show that the Eq. (\ref{hallcond}) gives $\sigma _{yx}=\frac{e^{2}}{h}%
n$, where $n\in 
\mathbb{Z}
$, we need to transform the integral of Eq. (\ref{hallcond}) to the integral
of the curvature of Berry connection in a closed loop. This necessitates a
'pull back' (i.e., undoing) the lattice transformation of Eq. (\ref{hallcond}%
), i.e., we revert to corresponding matrix elements.

\subsection{'Pull back' of the lattice Weyl transformation}

The pull-back process means we have to undo the lattice transformation of
SFLWT-NEGF transport equation, to return to its \textit{equivalent} matrix
element expressions. Consider the integrand in Eq. (\ref{hallcond}) given by
the partial derivatives of lattice Weyl transformed quantities.%
\begin{eqnarray}
&&\left[ \frac{\partial ^{\left( a\right) }}{\partial \mathcal{\vec{K}}_{x}}%
\frac{\partial ^{\left( b\right) }}{\partial \mathcal{\vec{K}}_{y}}-\frac{%
\partial ^{\left( a\right) }}{\partial \mathcal{\vec{K}}_{y}}\frac{\partial
^{\left( b\right) }}{\partial \mathcal{\vec{K}}_{x}}\right] H^{\left(
a\right) }\left( \mathcal{\vec{K}},\mathcal{E}\right) \left( -iG^{<\left(
b\right) }\left( \mathcal{\vec{K}},\mathcal{E}\right) \right)  \nonumber \\
&=&\left[ \frac{\partial H^{\left( a\right) }\left( \mathcal{\vec{K}},%
\mathcal{E}\right) }{\partial \mathcal{\vec{K}}_{x}}\frac{\partial
G^{<\left( b\right) }\left( \mathcal{\vec{K}},\mathcal{E}\right) }{\partial 
\mathcal{\vec{K}}_{y}}-\frac{\partial H^{\left( a\right) }\left( \mathcal{%
\vec{K}},\mathcal{E}\right) }{\partial \mathcal{\vec{K}}_{y}}\frac{\partial
G^{<\left( b\right) }\left( \mathcal{\vec{K}},\mathcal{E}\right) }{\partial 
\mathcal{\vec{K}}_{x}}\right] \text{.}  \label{kxky}
\end{eqnarray}%
The trick is to 'pull back' (undo) the lattice Weyl transformation to touch
base with Berry connection and Berry curvature. Take first the term of Eq. (%
\ref{kxky}), where,%
\begin{equation}
\frac{\partial H^{\left( a\right) }\left( \mathcal{\vec{K}},\mathcal{E}%
\right) }{\partial k_{x}}=\hbar \frac{\partial H^{\left( a\right) }\left( 
\mathcal{\vec{K}},\mathcal{E}\right) }{\partial \mathcal{\vec{K}}_{x}}\text{.%
}  \label{firsttrm}
\end{equation}%
From Eq. (\ref{sump}) this can be written as a lattice Weyl transform $%
\mathcal{W}$ in the form,%
\begin{eqnarray}
&&\frac{\partial H^{\left( a\right) }\left( \mathcal{\vec{K}},\mathcal{E}%
\right) }{\partial \mathcal{\vec{K}}_{x}}  \nonumber \\
&=&\mathcal{W}\left\{ \left( \frac{\partial }{\partial \mathcal{\vec{K}}%
_{x}^{\alpha }}+\frac{\partial }{\partial \mathcal{\vec{K}}_{x}^{^{\beta }}}%
\right) \left\langle \alpha ,\mathcal{\vec{K}},\mathcal{E}\right\vert \hat{H}%
\left\vert \beta ,\mathcal{\vec{K}},\mathcal{E}\right\rangle \right\} , 
\nonumber \\
&=&\mathcal{W}\left\{ \left\langle \alpha ,\frac{\partial }{\partial 
\mathcal{\vec{K}}_{x}}\mathcal{\vec{K}},\mathcal{E}\right\vert \hat{H}%
\left\vert \beta ,\mathcal{\vec{K}},\mathcal{E}\right\rangle +\left\langle
\alpha ,\mathcal{\vec{K}},\mathcal{E}\right\vert \hat{H}\left\vert \beta ,%
\frac{\partial }{\partial \mathcal{\vec{K}}_{x}}\mathcal{\vec{K}},\mathcal{E}%
\right\rangle \right\} ,  \nonumber \\
&=&\mathcal{W}\left\{ E_{\beta }\left( \mathcal{\vec{K}},\mathcal{E}\right)
\left\langle \alpha ,\frac{\partial }{\partial \mathcal{\vec{K}}_{x}}%
\mathcal{\vec{K}},\mathcal{E}\right\vert \left\vert \beta ,\mathcal{\vec{K}},%
\mathcal{E}\right\rangle +E_{\alpha }\left( \mathcal{\vec{K}},\mathcal{E}%
\right) \left\langle \alpha ,\mathcal{\vec{K}},\mathcal{E}\right\vert
\left\vert \beta ,\frac{\partial }{\partial \mathcal{\vec{K}}_{x}}\mathcal{%
\vec{K}},\mathcal{E}\right\rangle \right\} ,  \nonumber \\
&=&\mathcal{W}\left\{ \left( E_{\beta }\left( \mathcal{\vec{K}},\mathcal{E}%
\right) -E_{\alpha }\left( \mathcal{\vec{K}},\mathcal{E}\right) \right)
\left\langle \alpha ,\frac{\partial }{\partial \mathcal{\vec{K}}_{x}}%
\mathcal{\vec{K}},\mathcal{E}\right\vert \left\vert \beta ,\mathcal{\vec{K}},%
\mathcal{E}\right\rangle \right\} \text{,}  \label{LWTdhdKx}
\end{eqnarray}%
where we defined%
\[
\left\langle \alpha ,\frac{\partial }{\partial \mathcal{\vec{K}}_{x}}%
\mathcal{\vec{K}},\mathcal{E}\right\vert \equiv \frac{\partial }{\partial 
\mathcal{\vec{K}}_{x}}\left\langle \alpha ,\mathcal{\vec{K}},\mathcal{E}%
\right\vert . 
\]

We also have%
\[
\frac{\partial G^{<\left( b\right) }\left( \mathcal{\vec{K}},\mathcal{E}%
\right) }{\partial \mathcal{K}_{y}}=\mathcal{W}\left\{ \left( \frac{\partial 
}{\partial \mathcal{\vec{K}}_{y}^{\beta }}+\frac{\partial }{\partial 
\mathcal{\vec{K}}_{y}^{^{\alpha }}}\right) \left\langle \beta ,\mathcal{\vec{%
K}},\mathcal{E}\right\vert \left( i\hat{\rho}\right) \left\vert \alpha ,%
\mathcal{\vec{K}},\mathcal{E}\right\rangle \right\} \text{,} 
\]%
where $\hat{\rho}$ is the density matrix operator. From Eq. (\ref{timedepend}%
), we take the time dependence of 
\[
\left\langle \beta ,\mathcal{\vec{K}},\mathcal{E}\right\vert \left( i\hat{%
\rho}\right) \left\vert \alpha ,\mathcal{\vec{K}},\mathcal{E}\right\rangle 
\]
to be given by 
\[
i\left\langle \beta ,\mathcal{\vec{K}},\mathcal{E}\right\vert \hat{\rho}%
\left( 0\right) \left\vert \alpha ,\mathcal{\vec{K}},\mathcal{E}%
\right\rangle e^{i\omega _{\alpha \beta }t}. 
\]

We have\footnote{%
Here we use the definition of Green's function without the factor $\hbar $,
following traditional treatments, i.e. $\rho \left( 1,2\right)
=-iG^{<}\left( 1,2\right) $.}%
\[
\frac{\partial G^{<\left( b\right) }\left( \mathcal{\vec{K}},\mathcal{E}%
\right) }{\partial \mathcal{K}_{y}}=\mathcal{W}\left\{ 
\begin{array}{c}
\left\langle \beta ,\frac{\partial }{\partial \mathcal{\vec{K}}_{y}}\mathcal{%
\vec{K}},\mathcal{E}\right\vert \left( i\hat{\rho}_{0}\right) \left\vert
\alpha ,\mathcal{\vec{K}},\mathcal{E}\right\rangle \\ 
+\left\langle \beta ,\mathcal{\vec{K}},\mathcal{E}\right\vert i\hat{\rho}%
_{0}\left\vert \alpha ,\frac{\partial }{\partial \mathcal{\vec{K}}_{y}}%
\mathcal{\vec{K}},\mathcal{E}\right\rangle%
\end{array}%
\right\} e^{i\omega _{\alpha \beta }t}\text{.} 
\]%
The density matrix operator $\hat{\rho}_{0}$\ is of the form,%
\begin{eqnarray*}
\hat{\rho}_{o} &=&\sum\limits_{m}\rho _{m}\left\vert m\right\rangle
\left\langle m\right\vert \\
\hat{\rho}_{o}\left\vert m\right\rangle &=&\rho _{m}\left\vert
m\right\rangle =f\left( E_{m}\right) \left\vert m\right\rangle \\
\left\langle m\right\vert \hat{\rho}_{o}\left\vert n\right\rangle &=&\rho
_{mm}=f\left( E_{n}\right) \delta _{mn}\text{ or }f\left( E_{m}\right)
\delta _{mn}
\end{eqnarray*}%
where the weight function is the Fermi-Dirac function,%
\[
\rho _{0}^{m}=f\left( E_{m}\right) 
\]%
Hence%
\begin{eqnarray*}
i\hat{\rho}_{o}\left\vert \alpha ,\mathcal{\vec{K}},\mathcal{E}\right\rangle
&=&i\sum\limits_{\gamma }\left\vert \gamma ,\mathcal{\vec{K}},\mathcal{E}%
\right\rangle \rho _{0}^{\gamma }\left\langle \gamma ,\mathcal{\vec{K}},%
\mathcal{E}\right\vert \left\vert \alpha ,\mathcal{\vec{K}},\mathcal{E}%
\right\rangle \\
&=&i\left\vert \alpha ,\mathcal{\vec{K}},\mathcal{E}\right\rangle f\left(
E_{\alpha }\right) \text{.}
\end{eqnarray*}%
Similarly,%
\begin{eqnarray*}
i\left\langle \beta ,\mathcal{\vec{K}},\mathcal{E}\right\vert \left( \hat{%
\rho}_{0}\right) &=&i\left\langle \beta ,\mathcal{\vec{K}},\mathcal{E}%
\right\vert \sum\limits_{\gamma }\left\vert \gamma ,\mathcal{\vec{K}},%
\mathcal{E}\right\rangle \rho _{0}\left\langle \gamma ,\mathcal{\vec{K}},%
\mathcal{E}\right\vert \\
&=&if\left( E_{\beta }\right) \left\langle \beta ,\mathcal{\vec{K}},\mathcal{%
E}\right\vert \text{.}
\end{eqnarray*}%
Hence%
\[
\frac{\partial G^{<\left( b\right) }\left( \mathcal{\vec{K}},\mathcal{E}%
\right) }{\partial \mathcal{K}_{y}}=\mathcal{W}\left\{ 
\begin{array}{c}
i\left\langle \beta ,\frac{\partial }{\partial \mathcal{\vec{K}}_{y}}%
\mathcal{\vec{K}},\mathcal{E}\right\vert \left\vert \alpha ,\mathcal{\vec{K}}%
,\mathcal{E}\right\rangle \rho _{0}^{\alpha } \\ 
i\left\langle \beta ,\mathcal{\vec{K}},\mathcal{E}\right\vert \left\vert
\alpha ,\frac{\partial }{\partial \mathcal{\vec{K}}_{y}}\mathcal{\vec{K}},%
\mathcal{E}\right\rangle \rho _{0}^{\beta }%
\end{array}%
\right\} e^{i\omega _{\alpha \beta }t}\text{.} 
\]%
Shifting the first derivative to the right, we have%
\begin{eqnarray*}
\frac{\partial G^{<\left( b\right) }\left( \mathcal{\vec{K}},\mathcal{E}%
\right) }{\partial \mathcal{K}_{y}} &=&\mathcal{W}\left\{ 
\begin{array}{c}
-i\left\langle \beta ,\mathcal{\vec{K}},\mathcal{E}\right\vert \left\vert
\alpha ,\frac{\partial }{\partial \mathcal{\vec{K}}_{y}}\mathcal{\vec{K}},%
\mathcal{E}\right\rangle f\left( E_{\alpha }\right) \\ 
i\left\langle \beta ,\mathcal{\vec{K}},\mathcal{E}\right\vert \left\vert
\alpha ,\frac{\partial }{\partial \mathcal{\vec{K}}_{y}}\mathcal{\vec{K}},%
\mathcal{E}\right\rangle f\left( E_{\beta }\right)%
\end{array}%
\right\} e^{i\omega _{\alpha \beta }t} \\
&=&\mathcal{W}\left[ \left\{ i\left( f\left( E_{\beta }\right) -f\left(
E_{\alpha }\right) \right) \left\langle \beta ,\mathcal{\vec{K}},\mathcal{E}%
\right\vert \left\vert \alpha ,\frac{\partial }{\partial \mathcal{\vec{K}}%
_{y}}\mathcal{\vec{K}},\mathcal{E}\right\rangle \right\} e^{i\omega _{\alpha
\beta }t}\right] \text{.}
\end{eqnarray*}%
For energy scale it is convenient to chose $f\left( E_{\alpha }\right) $ in
the above equation, with the viewpoint that $\alpha $-state is far remove
from the $\beta $-state in gapped states, so that we can set $f\left(
E_{\beta }\right) \simeq 0$. The case $\alpha =\beta $ is indeterminate so
that by setting $f\left( E_{\beta }\right) \simeq 0$ renders the summation
to be well-defined, hence the need for gapped states. Therefore%
\begin{eqnarray*}
&&\frac{\partial H\left( \mathcal{\vec{K}},\mathcal{E}\right) }{\partial 
\mathcal{K}_{x}}\frac{\partial G^{<}\left( \mathcal{\vec{K}},\mathcal{E}%
\right) }{\partial \mathcal{K}_{y}} \\
&=&\left\{ 
\begin{array}{c}
\mathcal{W}\left[ \left( -\left( E_{\beta }\left( \mathcal{\vec{K}},\mathcal{%
E}\right) -E_{\alpha }\left( \mathcal{\vec{K}},\mathcal{E}\right) \right)
\right) \left\{ \left\langle \alpha ,\frac{\partial }{\partial \mathcal{\vec{%
K}}_{x}^{^{^{\prime \prime }}}}\mathcal{\vec{K}},\mathcal{E}\right\vert
\left\vert \beta ,\mathcal{\vec{K}},\mathcal{E}\right\rangle \right\} \right]
\\ 
\times \mathcal{W}\left[ \left\{ -if\left( E_{\alpha }\right) \left\langle
\beta ,\mathcal{\vec{K}},\mathcal{E}\right\vert \left\vert \alpha ,\frac{%
\partial }{\partial \mathcal{\vec{K}}_{y}}\mathcal{\vec{K}},\mathcal{E}%
\right\rangle \right\} e^{i\omega _{\alpha \beta }t}\right]%
\end{array}%
\right\} , \\
&=&\left\{ 
\begin{array}{c}
\mathcal{W}\left[ \left( E_{\beta }\left( \mathcal{\vec{K}},\mathcal{E}%
\right) -E_{\alpha }\left( \mathcal{\vec{K}},\mathcal{E}\right) \right)
\left\{ \left\langle \alpha ,\frac{\partial }{\partial \mathcal{\vec{K}}%
_{x}^{^{^{\prime \prime }}}}\mathcal{\vec{K}},\mathcal{E}\right\vert
\left\vert \beta ,\mathcal{\vec{K}},\mathcal{E}\right\rangle \right\} \right]
\\ 
\times \mathcal{W}\left[ \left\{ i\left\langle \beta ,\mathcal{\vec{K}},%
\mathcal{E}\right\vert \left\vert \alpha ,\frac{\partial }{\partial \mathcal{%
\vec{K}}_{y}}\mathcal{\vec{K}},\mathcal{E}\right\rangle \right\} \right]
f\left( E_{\alpha }\right) e^{i\omega _{\alpha \beta }t}%
\end{array}%
\right\} \text{.}
\end{eqnarray*}%
Since it appears as a product of two Weyl transforms, it must be a trace
formula in the \textit{untransformed or pulled back} version, i.e., for the
remaing indices $\alpha $ and $\beta $ we must be a summation, 
\begin{eqnarray*}
&&\frac{\partial H\left( \mathcal{\vec{K}},\mathcal{E}\right) }{\partial 
\mathcal{K}_{x}}\frac{\partial G^{<}\left( \mathcal{\vec{K}},\mathcal{E}%
\right) }{\partial \mathcal{K}_{y}} \\
&=&\mathcal{W}\left[ 
\begin{array}{c}
\sum\limits_{\alpha ,\beta }\left\{ 
\begin{array}{c}
\left( E_{\beta }\left( \mathcal{\vec{K}},\mathcal{E}\right) -E_{\alpha
}\left( \mathcal{\vec{K}},\mathcal{E}\right) \right) \\ 
\times \left\{ \left\langle \alpha ,\frac{\partial }{\partial \mathcal{K}_{x}%
}\mathcal{\vec{K}},\mathcal{E}\right\vert \left\vert \beta ,\mathcal{\vec{K}}%
,\mathcal{E}\right\rangle \right\} \left\{ \left\langle \beta ,\mathcal{\vec{%
K}},\mathcal{E}\right\vert \left\vert \alpha ,\frac{\partial }{\partial 
\mathcal{K}_{y}}\mathcal{\vec{K}},\mathcal{E}\right\rangle \right\}
e^{i\omega _{\alpha \beta }t}%
\end{array}%
\right\} \\ 
\times \ i\left( f\left( E_{\alpha }\right) \right) .%
\end{array}%
\right] \text{.}
\end{eqnarray*}%
Similarly, we have%
\begin{eqnarray*}
&&\frac{\partial H\left( \mathcal{\vec{K}},\mathcal{E}\right) }{\partial 
\mathcal{K}_{y}}\frac{\partial G^{<}\left( \mathcal{\vec{K}},\mathcal{E}%
\right) }{\partial \mathcal{K}_{x}} \\
&=&\mathcal{W}\left[ 
\begin{array}{c}
\sum\limits_{\alpha ,\beta }\left\{ 
\begin{array}{c}
\left( E_{\beta }\left( \mathcal{\vec{K}},\mathcal{E}\right) -E_{\alpha
}\left( \mathcal{\vec{K}},\mathcal{E}\right) \right) \\ 
\times \left\{ \left\langle \alpha ,\frac{\partial }{\partial \mathcal{K}_{y}%
}\mathcal{\vec{K}},\mathcal{E}\right\vert \left\vert \beta ,\mathcal{\vec{K}}%
,\mathcal{E}\right\rangle \right\} \left\{ \left\langle \beta ,\mathcal{\vec{%
K}},\mathcal{E}\right\vert \left\vert \alpha ,\frac{\partial }{\partial 
\mathcal{K}_{x}}\mathcal{\vec{K}},\mathcal{E}\right\rangle \right\}
e^{i\omega _{\alpha \beta }t}%
\end{array}%
\right\} \\ 
\times \ if\left( E_{\alpha }\right) .%
\end{array}%
\right] \text{.}
\end{eqnarray*}

Therefore we obtain,%
\begin{eqnarray*}
&&\left[ \frac{\partial H\left( \mathcal{\vec{K}},\mathcal{E}\right) }{%
\partial \mathcal{K}_{x}}\frac{\partial G^{<}\left( \mathcal{\vec{K}},%
\mathcal{E}\right) }{\partial \mathcal{K}_{y}}-\frac{\partial H\left( 
\mathcal{\vec{K}},\mathcal{E}\right) }{\partial \mathcal{K}_{y}}\frac{%
\partial G^{<}\left( \mathcal{\vec{K}},\mathcal{E}\right) }{\partial 
\mathcal{K}_{x}}\right]  \\
&=&\mathcal{W}\left[ 
\begin{array}{c}
\sum\limits_{\alpha ,\beta }\left\{ 
\begin{array}{c}
\left( E_{\beta }\left( \mathcal{\vec{K}},\mathcal{E}\right) -E_{\alpha
}\left( \mathcal{\vec{K}},\mathcal{E}\right) \right)  \\ 
\times \left[ 
\begin{array}{c}
\left\{ \left\langle \alpha ,\frac{\partial }{\partial \mathcal{K}_{x}}%
\mathcal{\vec{K}},\mathcal{E}\right\vert \left\vert \beta ,\mathcal{\vec{K}},%
\mathcal{E}\right\rangle \right\} \left\{ \left\langle \beta ,\mathcal{\vec{K%
}},\mathcal{E}\right\vert \left\vert \alpha ,\frac{\partial }{\partial 
\mathcal{K}_{y}}\mathcal{\vec{K}},\mathcal{E}\right\rangle \right\}  \\ 
-\left\{ \left\langle \alpha ,\frac{\partial }{\partial \mathcal{K}_{y}}%
\mathcal{\vec{K}},\mathcal{E}\right\vert \left\vert \beta ,\mathcal{\vec{K}},%
\mathcal{E}\right\rangle \right\} \left\{ \left\langle \beta ,\mathcal{\vec{K%
}},\mathcal{E}\right\vert \left\vert \alpha ,\frac{\partial }{\partial 
\mathcal{K}_{x}}\mathcal{\vec{K}},\mathcal{E}\right\rangle \right\} 
\end{array}%
\right] 
\end{array}%
\right\}  \\ 
\times \ ie^{i\omega _{\alpha \beta }t}\left( f\left( E_{\alpha }\right)
\right) 
\end{array}%
\right] \text{.}
\end{eqnarray*}%
Now the LHS of Eq. (\ref{eqKE}), namely%
\begin{eqnarray}
&&\left( \frac{a}{\left( 2\pi \hbar \right) }\right) ^{2}\int d\mathcal{\vec{%
K}}_{x}d\mathcal{\vec{K}}_{y}\frac{e}{a^{2}}\frac{\partial \mathcal{E}}{%
\partial \mathcal{\vec{K}}_{y}}G^{<}\left( \mathcal{\vec{K}},\mathcal{E}%
\right)   \nonumber \\
&=&\left( \frac{a}{\left( 2\pi \hbar \right) }\right) ^{2}\int d\mathcal{%
\vec{K}}_{x}d\mathcal{\vec{K}}_{y}\frac{e}{a^{2}}\frac{\partial H}{\partial 
\mathcal{\vec{K}}_{y}}G^{<}\left( \mathcal{\vec{K}},\mathcal{E}\right) \text{%
.}  \label{LHS}
\end{eqnarray}%
Using the result of Eq. (\ref{LWTdhdKx}), we have%
\begin{eqnarray*}
&&\frac{\partial H\left( \mathcal{\vec{K}},\mathcal{E}\right) }{\partial 
\mathcal{\vec{K}}_{y}} \\
&=&\mathcal{W}\left\{ \left[ \left( E_{\alpha }\left( \mathcal{\vec{K}},%
\mathcal{E}\right) -E_{\beta }\left( \mathcal{\vec{K}},\mathcal{E}\right)
\right) \right] \left\langle \alpha ,\mathcal{\vec{K}},\mathcal{E}%
\right\vert \left\vert \beta ,\frac{\partial }{\partial \mathcal{\vec{K}}%
_{y}^{^{^{\prime \prime }}}}\mathcal{\vec{K}},\mathcal{E}\right\rangle
\right\} , \\
&=&\mathcal{W}\left\{ \left[ \left( E_{\beta }\left( \mathcal{\vec{K}},%
\mathcal{E}\right) -E_{\alpha }\left( \mathcal{\vec{K}},\mathcal{E}\right)
\right) \right] \left\langle \alpha ,\frac{\partial }{\partial \mathcal{\vec{%
K}}_{y}}\mathcal{\vec{K}},\mathcal{E}\right\vert \left\vert \beta ,\mathcal{%
\vec{K}},\mathcal{E}\right\rangle \right\} , \\
&=&\mathcal{W}\left\{ \omega _{\beta \alpha }\left\langle \alpha ,\frac{%
\partial }{\partial k_{y}}\mathcal{\vec{K}},\mathcal{E}\right\vert
\left\vert \beta ,\mathcal{\vec{K}},\mathcal{E}\right\rangle \right\}
=\left\langle \alpha ,\mathcal{\vec{K}},\mathcal{E}\right\vert
v_{g,y}\left\vert \beta ,\mathcal{\vec{K}},\mathcal{E}\right\rangle ,
\end{eqnarray*}%
where%
\[
\omega _{\beta \alpha }\left\langle \alpha ,\nabla _{\mathcal{\vec{K}}}\vec{p%
}\right\vert \left\vert \beta ,\vec{p}\right\rangle =\left\langle \alpha ,%
\vec{p}\right\vert \vec{v}_{g}\left\vert \beta ,\vec{p}\right\rangle 
\]%
Likewise%
\[
G^{<}\left( \mathcal{\vec{K}},\mathcal{E}\right) =i\mathcal{W}\left(
\left\langle \beta ,\mathcal{\vec{K}},\mathcal{E}\right\vert \hat{\rho}%
\left\vert \alpha ,\mathcal{\vec{K}},\mathcal{E}\right\rangle \right) 
\]%
Again, since Eq. (\ref{LHS}) is a product of lattice Weyl transform, it must
be a trace in the \textit{untransformed} version, i.e., 
\begin{eqnarray*}
&&\left( \frac{a}{\left( 2\pi \hbar \right) }\right) ^{2}\int d\mathcal{\vec{%
K}}_{x}d\mathcal{\vec{K}}_{y}\frac{e}{a^{2}}\frac{\partial H}{\partial 
\mathcal{\vec{K}}_{y}}G^{<}\left( \mathcal{\vec{K}},\mathcal{E}\right)  \\
&=&\mathcal{W}\left\{ 
\begin{array}{c}
i\int \left( \frac{a}{\left( 2\pi \hbar \right) }\right) ^{2}d\mathcal{\vec{K%
}}_{x}d\mathcal{\vec{K}}_{y} \\ 
\times \sum\limits_{\alpha ,\beta }\left\langle \alpha ,\mathcal{\vec{K}},%
\mathcal{E}\right\vert \frac{e}{a^{2}}v_{y}\left\vert \beta ,\mathcal{\vec{K}%
},\mathcal{E}\right\rangle \left\langle \beta ,\mathcal{\vec{K}},\mathcal{E}%
\right\vert \hat{\rho}\left\vert \alpha ,\mathcal{\vec{K}},\mathcal{E}%
\right\rangle 
\end{array}%
\right\} , \\
&=&\mathcal{W}\left\{ iTr\left( \frac{e}{a^{2}}\hat{v}_{g,y}\right) \hat{\rho%
}\right\} =i\mathcal{W}\left\{ Tr\left( \hat{\jmath}_{y}\hat{\rho}\right)
\right\} =i\mathcal{W}\left\{ Tr\left( \hat{\jmath}_{y}\ \hat{\rho}\right)
\right\} , \\
&=&i\mathcal{W}\left\{ \left\langle \hat{\jmath}_{y}\left( t\right)
\right\rangle \right\} \text{.}
\end{eqnarray*}%
For calculating the conductivity we are interested in the term multiplying
the first-order in electric field. We can now convert the quantum transport
equation in the transformed space, Eq. (\ref{eqKE}), 
\begin{eqnarray*}
&&\left( \frac{a}{\left( 2\pi \hbar \right) }\right) ^{2}\int d\mathcal{\vec{%
K}}_{x}d\mathcal{\vec{K}}_{y}\frac{e}{a^{2}}\frac{\partial \mathcal{E}}{%
\partial \mathcal{\vec{K}}_{y}}\left[ -iG^{<}\left( \mathcal{\vec{K}},%
\mathcal{E}\right) \right]  \\
&=&e^{2}\left\vert \vec{F}\right\vert \frac{1}{\left( 2\pi \hbar \right) ^{2}%
}\int \int \int d\mathcal{\vec{K}}_{x}d\mathcal{\vec{K}}_{y}dt\ \left[ \frac{%
\partial ^{\left( a\right) }}{\partial \mathcal{\vec{K}}_{x}}\frac{\partial
^{\left( b\right) }}{\partial \mathcal{\vec{K}}_{y}}-\frac{\partial ^{\left(
a\right) }}{\partial \mathcal{\vec{K}}_{y}}\frac{\partial ^{\left( b\right) }%
}{\partial \mathcal{\vec{K}}_{x}}\right]  \\
&&\times H^{\left( a\right) }\left( \mathcal{\vec{K}},\mathcal{E}\right)
\left( -iG^{<\left( b\right) }\left( \mathcal{\vec{K}},\mathcal{E}\right)
\right) \text{,}
\end{eqnarray*}%
to the equivalent matrix element expressions by undoing or 'pulling back'
the lattice Weyl transformation $\mathcal{W}$, which amounts to canceling $%
\mathcal{W}$ in both side of the equation given by, 
\begin{eqnarray}
&&\mathcal{W}\left\{ \left\langle \hat{\jmath}_{y}\left( t\right)
\right\rangle \right\}   \nonumber \\
&=&\mathcal{W}\left\{ 
\begin{array}{c}
\frac{e^{2}}{h}\left\vert \vec{F}\right\vert \frac{1}{\left( 2\pi \hbar
\right) }\int \int \int d\mathcal{\vec{K}}_{x}d\mathcal{\vec{K}}_{y}dt \\ 
\times \sum\limits_{\alpha ,\beta }\left[ 
\begin{array}{c}
\left( E_{\beta }\left( \mathcal{\vec{K}},\mathcal{E}\right) -E_{\alpha
}\left( \mathcal{\vec{K}},\mathcal{E}\right) \right)  \\ 
\times \left\{ 
\begin{array}{c}
\left\langle \alpha ,\frac{\partial }{\partial \mathcal{\vec{K}}_{x}}%
\mathcal{\vec{K}},\mathcal{E}\right\vert \left\vert \beta ,\mathcal{\vec{K}},%
\mathcal{E}\right\rangle \left\langle \beta ,\mathcal{\vec{K}},\mathcal{E}%
\right\vert \left\vert \alpha ,\frac{\partial }{\partial \mathcal{\vec{K}}%
_{y}}\mathcal{\vec{K}},\mathcal{E}\right\rangle  \\ 
-\left\langle \alpha ,\frac{\partial }{\partial \mathcal{\vec{K}}_{y}}%
\mathcal{\vec{K}},\mathcal{E}\right\vert \left\vert \beta ,\mathcal{\vec{K}},%
\mathcal{E}\right\rangle \left\langle \beta ,\mathcal{\vec{K}},\mathcal{E}%
\right\vert \left\vert \alpha ,\frac{\partial }{\partial \mathcal{\vec{K}}%
_{x}}\mathcal{\vec{K}},\mathcal{E}\right\rangle 
\end{array}%
\right\}  \\ 
\times \ e^{i\omega _{\alpha \beta }t}f\left( E_{\alpha }\right) 
\end{array}%
\right] \text{.}%
\end{array}%
\right\}   \nonumber \\
&&  \label{pullback}
\end{eqnarray}%
The time integral of the RHS amounts to taking zero-order time dependence
[zero electric field] of the rest of the integrand, then we have for the
remaining time-dependence, explicitly integrated as, 
\begin{eqnarray*}
\int\limits_{-\infty }^{0}dt\exp i\omega _{\alpha \beta }t &=&\left. \frac{%
\exp \exp i\omega _{\alpha \beta }t}{i\omega _{\alpha \beta }}\right\vert
_{\tau =-\infty }^{\tau =0} \\
&=&\left. \frac{\exp \left( i\left( \omega _{\alpha \beta }-i\eta \right)
\tau \right) }{i\omega _{\alpha \beta }}\right\vert _{\tau =-\infty }^{\tau
=0}=\frac{1}{i\omega _{\alpha \beta }}
\end{eqnarray*}%
Thus eliminating the time integral we finally obtain. 
\begin{eqnarray*}
&&\left\langle \hat{\jmath}_{y}\left( t\right) \right\rangle  \\
&=&-i\frac{e^{2}}{h}\left\vert \vec{F}\right\vert \frac{1}{\left( 2\pi \hbar
\right) }\int \int d\mathcal{\vec{K}}_{x}d\mathcal{\vec{K}}_{y} \\
&&\times \sum\limits_{\alpha ,\beta }\left[ 
\begin{array}{c}
f\left( E_{\alpha }\right) \left( -\frac{\hbar \omega _{\alpha \beta }}{%
\omega _{\alpha \beta }}\right)  \\ 
\times \left\{ 
\begin{array}{c}
\left\langle \alpha ,\frac{\partial }{\partial \mathcal{\vec{K}}_{x}}%
\mathcal{\vec{K}},\mathcal{E}\right\vert \left\vert \beta ,\mathcal{\vec{K}},%
\mathcal{E}\right\rangle \left\langle \beta ,\mathcal{\vec{K}},\mathcal{E}%
\right\vert \left\vert \alpha ,\frac{\partial }{\partial \mathcal{\vec{K}}%
_{y}}\mathcal{\vec{K}},\mathcal{E}\right\rangle  \\ 
-\left\langle \alpha ,\frac{\partial }{\partial \mathcal{\vec{K}}_{y}}%
\mathcal{\vec{K}},\mathcal{E}\right\vert \left\vert \beta ,\mathcal{\vec{K}},%
\mathcal{E}\right\rangle \left\langle \beta ,\mathcal{\vec{K}},\mathcal{E}%
\right\vert \left\vert \alpha ,\frac{\partial }{\partial \mathcal{\vec{K}}%
_{x}}\mathcal{\vec{K}},\mathcal{E}\right\rangle 
\end{array}%
\right\} 
\end{array}%
\right] 
\end{eqnarray*}%
which reduces to%
\begin{eqnarray}
&&\left\langle \hat{\jmath}_{y}\left( t\right) \right\rangle   \nonumber \\
&=&i\frac{e^{2}}{h}\left\vert \vec{F}\right\vert \frac{1}{\left( 2\pi
\right) }\int \int dk_{x}dk_{y}  \nonumber \\
&&\times \sum\limits_{\alpha ,\beta }f\left( E_{\alpha }\right) \left\{ 
\begin{array}{c}
\left\langle \alpha ,\frac{\partial }{\partial k_{x}}\mathcal{\vec{K}},%
\mathcal{E}\right\vert \left\vert \beta ,\mathcal{\vec{K}},\mathcal{E}%
\right\rangle \left\langle \beta ,\mathcal{\vec{K}},\mathcal{E}\right\vert
\left\vert \alpha ,\frac{\partial }{\partial k_{y}}\mathcal{\vec{K}},%
\mathcal{E}\right\rangle  \\ 
-\left\langle \alpha ,\frac{\partial }{\partial k_{y}}\mathcal{\vec{K}},%
\mathcal{E}\right\vert \left\vert \beta ,\mathcal{\vec{K}},\mathcal{E}%
\right\rangle \left\langle \beta ,\mathcal{\vec{K}},\mathcal{E}\right\vert
\left\vert \alpha ,\frac{\partial }{\partial k_{x}}\mathcal{\vec{K}},%
\mathcal{E}\right\rangle 
\end{array}%
\right\}   \label{notint}
\end{eqnarray}%
Taking the Fourier transform of both sides, we obtain%
\begin{eqnarray}
&&\left\langle \hat{\jmath}_{y}\left( \omega \right) \right\rangle  
\nonumber \\
&=&i\frac{e^{2}}{h}\left\vert \vec{F}\right\vert \frac{\delta \left( \omega
\right) }{\left( 2\pi \right) }\int \int dk_{x}dk_{y}  \nonumber \\
&&\times \sum\limits_{\alpha ,\beta }\left\{ 
\begin{array}{c}
\left\langle \alpha ,\frac{\partial }{\partial k_{x}}\mathcal{\vec{K}},%
\mathcal{E}\right\vert \left\vert \beta ,\mathcal{\vec{K}},\mathcal{E}%
\right\rangle \left\langle \beta ,\mathcal{\vec{K}},\mathcal{E}\right\vert
\left\vert \alpha ,\frac{\partial }{\partial k_{y}}\mathcal{\vec{K}},%
\mathcal{E}\right\rangle  \\ 
-\left\langle \alpha ,\frac{\partial }{\partial k_{y}}\mathcal{\vec{K}},%
\mathcal{E}\right\vert \left\vert \beta ,\mathcal{\vec{K}},\mathcal{E}%
\right\rangle \left\langle \beta ,\mathcal{\vec{K}},\mathcal{E}\right\vert
\left\vert \alpha ,\frac{\partial }{\partial k_{x}}\mathcal{\vec{K}},%
\mathcal{E}\right\rangle 
\end{array}%
\right\} f\left( E_{\alpha }\right) .  \label{currenteq}
\end{eqnarray}%
Taking the limit $\omega \Longrightarrow 0$ and summing over the states $%
\beta $, we readily obtain the conductivity, $\sigma _{yx}$. 
\begin{equation}
\sigma _{yx}=\frac{e^{2}}{h}\sum\limits_{\alpha }f\left( E_{\alpha }\right) 
\frac{i}{\left( 2\pi \right) }\int \int dk_{x}dk_{y}\left[ 
\begin{array}{c}
\left\langle \alpha ,\frac{\partial }{\partial k_{x}}\mathcal{\vec{K}},%
\mathcal{E}\right\vert \left\vert \alpha ,\frac{\partial }{\partial k_{y}}%
\mathcal{\vec{K}},\mathcal{E}\right\rangle  \\ 
-\left\langle \alpha ,\frac{\partial }{\partial k_{y}}\mathcal{\vec{K}},%
\mathcal{E}\right\vert \left\vert \alpha ,\frac{\partial }{\partial k_{x}}%
\mathcal{\vec{K}},\mathcal{E}\right\rangle 
\end{array}%
\right] \text{,}  \label{hall}
\end{equation}%
where 
\[
\mathcal{\vec{K}}={\vec{p}}+e\vec{F}t\ .
\]%
Note the transformation from $\mathcal{\vec{K}}$ $\Longrightarrow $ $\vec{p}$
in the integration in both Eqs. (\ref{currenteq}) and (\ref{hall}) has a
Jacobian unity. Equation (\ref{hall}) is the same expression that can
obtained to derive the integer quantum Hall effect from Kubo formula \cite%
{tknn}.

We now prove that for each statevector, $\left\vert \alpha ,\vec{k}%
\right\rangle $, the expression, 
\begin{equation}
\frac{i}{\left( 2\pi \right) }\int \int dk_{x}dk_{y}\ f\left( E_{\alpha
}\left( \vec{k}\right) \right) \left[ \left\langle \frac{\partial }{\partial
k_{x}}\alpha ,\vec{k}\right\vert \frac{\partial }{\partial k_{y}}\left\vert
\alpha ,\vec{k}\right\rangle -\left\langle \frac{\partial }{\partial k_{y}}%
\alpha ,\vec{k}\right\vert \frac{\partial }{\partial k_{x}}\left\vert \alpha
,\vec{k}\right\rangle \right] \text{,}  \label{chernNum}
\end{equation}%
is the winding number around the occupied contour in the Brillouin zone.
First we can rewrite the terms within the square bracket as%
\begin{eqnarray}
&&\left[ \left\langle \frac{\partial }{\partial k_{x}}\alpha ,\vec{k}%
\right\vert \frac{\partial }{\partial k_{y}}\left\vert \alpha ,\vec{k}%
\right\rangle -\left\langle \frac{\partial }{\partial k_{y}}\alpha ,\vec{k}%
\right\vert \frac{\partial }{\partial k_{x}}\left\vert \alpha ,\vec{k}%
\right\rangle \right]  \nonumber \\
&=&\left\langle \frac{\partial }{\partial \vec{k}}\alpha ,\vec{k}\right\vert
\times \frac{\partial }{\partial \vec{k}}\left\vert \alpha ,\vec{k}%
\right\rangle =\nabla _{\vec{k}}\times \left\langle \alpha ,\vec{k}%
\right\vert \frac{\partial }{\partial \vec{k}}\left\vert \alpha ,\vec{k}%
\right\rangle \text{.}  \label{curlA}
\end{eqnarray}%
The last term indicates the operation of the curl of the Berry connection
which is related to the quantization of Hall conductivity. This quantization
is due to the uniqueness of the parallel-transported wavefunction, which may
also have bearing on the self-consistent Bohr-Sommerfeld quantization \cite%
{fego}. To understand this, first we discuss how the phase of the
wavefunction relates to the Berry connection and Berry curvature. This is
given in the Appendix B.

At low temperature, we can just write Eq. (\ref{hall}) as,%
\begin{eqnarray}
\ \sigma _{yx} &=&\frac{ie^{2}}{2\pi \hbar }\frac{1}{\left( 2\pi \right) }%
\sum\limits_{\alpha }\int \int_{occupiedBZ}dk_{x}dk_{y}\ \left[ \nabla _{%
\vec{k}}\times \left\langle \alpha ,\vec{k}\right\vert \frac{\partial }{%
\partial \vec{k}}\left\vert \alpha ,\vec{k}\right\rangle \right] _{plane}, 
\nonumber \\
&=&\frac{e^{2}}{2\pi \hbar }\frac{i}{\left( 2\pi \right) }%
\sum\limits_{\alpha }\doint dk_{c}\ \left[ \left\langle \alpha ,\vec{k}%
\right\vert \frac{\partial }{\partial k_{c}}\left\vert \alpha ,\vec{k}%
\right\rangle \right] _{contour}\text{.}  \label{lowT}
\end{eqnarray}%
Now 
\begin{eqnarray}
i\Delta \phi _{total} &=&-\doint dk_{c}\ \left[ \left\langle \alpha ,\vec{k}%
\right\vert \frac{\partial }{\partial k_{c}}\left\vert \alpha ,\vec{k}%
\right\rangle \right] _{contour},  \nonumber \\
\Delta \phi _{total} &=&i\doint dk_{c}\ \left[ \left\langle \alpha ,\vec{k}%
\right\vert \frac{\partial }{\partial k_{c}}\left\vert \alpha ,\vec{k}%
\right\rangle \right] _{contour},  \nonumber \\
\frac{\Delta \phi _{total}}{2\pi } &=&\frac{i}{2\pi }\doint dk_{c}\ \left[
\left\langle \alpha ,\vec{k}\right\vert \frac{\partial }{\partial k_{c}}%
\left\vert \alpha ,\vec{k}\right\rangle \right] _{contour}\text{ }=n_{\alpha
}.  \label{phasechange}
\end{eqnarray}%
where $n_{\alpha }\ \epsilon \ 
\mathbb{Z}
$ is the winding number or the Chern number. Therefore%
\begin{eqnarray}
\ \sigma _{yx} &=&\frac{ie^{2}}{2\pi \hbar }\frac{1}{\left( 2\pi \right) }%
\sum\limits_{\alpha }\int \int_{occupiedBZ}dk_{x}dk_{y}\ \left[ \nabla _{%
\vec{k}}\times \left\langle \alpha ,\vec{k}\right\vert \frac{\partial }{%
\partial \vec{k}}\left\vert \alpha ,\vec{k}\right\rangle \right] _{plane}, 
\nonumber \\
&=&\frac{e^{2}}{2\pi \hbar }\frac{i}{\left( 2\pi \right) }%
\sum\limits_{\alpha }\doint dk_{c}\ \left[ \left\langle \alpha ,\vec{k}%
\right\vert \frac{\partial }{\partial k_{c}}\left\vert \alpha ,\vec{k}%
\right\rangle \right] _{contour\ or\ Wilson\ loop\ over\ BZ}\ ,  \nonumber \\
&=&\frac{e^{2}}{h}\sum\limits_{\alpha }\frac{\Delta \phi _{total}}{2\pi }%
=\sum\limits_{\alpha }\frac{e^{2}}{h}n_{\alpha }\text{,}  \label{bingo}
\end{eqnarray}%
over all occupied bands $\alpha ,$ where $n_{\alpha }\in 
\mathbb{Z}
$ is the topological first Chern (or \textit{winding}) number for each band.
Thus the the Hall conductivity is quantized in units of $\frac{e^{2}}{h}$ as
derive from Eq. (\ref{hallcond}) of the new method used here.

To touch base with a time-dependent perturbation of the Kubo current-current
correlation(KCCC) used by TKNN, we recall that in this particular approach,
a time varying electric field is indirectly used. We give a derivation of
the KCCC formula from our new quantum transport formalism in Appendix C.

\subsection{Berry curvature and orbital magnetic moment of 2-D systems}

Note that the orbital magnetic moment is given by,%
\begin{eqnarray*}
\left\langle \vec{M}\right\rangle &=&Tr\rho \frac{e}{2m}\vec{L}=Tr\rho \frac{%
e}{2m}\vec{Q}\times \vec{P}, \\
&=&-i\frac{e}{2m}\sum\limits_{\alpha ,p;\beta ,p}f\left( E_{\alpha }\left( 
\vec{p}\right) \right) \left\langle \alpha ,\vec{p}\right\vert \nabla _{\vec{%
k}}\left\vert \beta ,\vec{p}\right\rangle \times \left\langle \beta ,\vec{p}%
\right\vert \vec{P}\left\vert \alpha ,\vec{p}\right\rangle , \\
&=&-i\frac{e}{2}\sum\limits_{\alpha ,p;\beta ,p}f\left( E_{\alpha }\left( 
\vec{p}\right) \right) \left\langle \alpha ,\vec{p}\right\vert \nabla _{\vec{%
k}}\left\vert \beta ,\vec{p}\right\rangle \times \left\langle \beta ,\vec{p}%
\right\vert \vec{v}_{g}\left\vert \alpha ,\vec{p}\right\rangle .
\end{eqnarray*}%
Using Eq. (\ref{transv2}), we obtain%
\[
\left\langle \vec{M}\right\rangle =-i\frac{e}{2}\sum\limits_{\alpha ,p;\beta
,p}f\left( E_{\alpha }\left( \vec{p}\right) \right) \omega _{\beta \alpha
}\left\langle \alpha ,\vec{p}\right\vert \nabla _{\vec{k}}\left\vert \beta ,%
\vec{p}\right\rangle \times \left\langle \beta ,\vec{p}\right\vert \nabla _{%
\vec{k}}\left\vert \alpha ,\vec{p}\right\rangle \text{,} 
\]%
or%
\[
\left\langle \vec{M}\right\rangle =-i\frac{e}{2}\sum\limits_{\alpha ,p;\beta
,p}\frac{f\left( E_{\alpha }\left( \vec{p}\right) \right) }{\omega _{\alpha
\beta }}\left\langle \alpha ,\vec{p}\right\vert \vec{v}\left\vert \beta ,%
\vec{p}\right\rangle \times \left\langle \beta ,\vec{p}\right\vert \vec{v}%
\left\vert \alpha ,\vec{p}\right\rangle \text{.} 
\]%
So Berry's curvature implies the presence of orbital magnetic moment. 
\footnote{%
Using Eq. (\ref{transv2}), we obtain as written in some literature%
\begin{eqnarray*}
&&\left\langle M\right\rangle \\
&=&-\frac{e}{2}\left( i\right) ^{2}\frac{\hbar ^{2}}{m^{2}}%
\sum\limits_{\alpha ,p;\beta ,p}f\left( E_{\alpha }\left( \vec{p}\right)
\right) \frac{\left( E_{\alpha }\left( \vec{p}\right) -E_{\beta }\left( \vec{%
p}\right) \right) }{i\hbar }\frac{\left\langle \alpha ,\vec{p}\right\vert m%
\vec{v}\left\vert \beta ,\vec{p}\right\rangle }{\left( E_{\alpha }\left( 
\vec{p}\right) -E_{\beta }\left( \vec{p}\right) \right) }\times \frac{%
\left\langle \beta ,\vec{p}\right\vert m\vec{v}\left\vert \alpha ,\vec{p}%
\right\rangle }{\left( E_{\alpha }\left( \vec{p}\right) -E_{\beta }\left( 
\vec{p}\right) \right) } \\
&=&-i\frac{e\hbar }{2m^{2}}\sum\limits_{\alpha ,p;\beta ,p}f\left( E_{\alpha
}\left( \vec{p}\right) \right) \frac{\left\langle \alpha ,\vec{p}\right\vert
m\vec{v}\left\vert \beta ,\vec{p}\right\rangle \times \left\langle \beta ,%
\vec{p}\right\vert m\vec{v}\left\vert \alpha ,\vec{p}\right\rangle }{\left(
E_{\alpha }\left( \vec{p}\right) -E_{\beta }\left( \vec{p}\right) \right) }%
\text{.}
\end{eqnarray*}%
}

In fact we can pull out the Berry curvature by rewriting in the Heisenberg
picture,%
\begin{eqnarray*}
\left\langle \vec{M}\right\rangle &=&Tr\rho \frac{e}{2m}L=Tr\rho \frac{e}{2m}%
\vec{Q}\times \vec{P}, \\
&=&\frac{e}{2m}\sum\limits_{\alpha ,p;\beta ,p}f\left( E_{\alpha }\left( 
\vec{p}\right) \right) \left\langle \alpha ,\vec{p}\right\vert \vec{Q}%
\left\vert \beta ,\vec{p}\right\rangle \times \left\langle \beta ,\vec{p}%
\right\vert \vec{P}\left\vert \alpha ,\vec{p}\right\rangle , \\
&=&\frac{e}{2}\sum\limits_{\alpha ,p;\beta ,p}f\left( E_{\alpha }\left( \vec{%
p}\right) \right) \left\langle \alpha ,\vec{p}\right\vert e^{\frac{i}{\hbar }%
Ht}\vec{Q}_{S}e^{-\frac{i}{\hbar }Ht}\left\vert \beta ,\vec{p}\right\rangle
\times \left\langle \beta ,\vec{p}\right\vert \vec{v}\left\vert \alpha ,\vec{%
p}\right\rangle , \\
&=&\frac{e}{2}\sum\limits_{\alpha ,p;\beta ,p}f\left( E_{\alpha }\left( \vec{%
p}\right) \right) e^{i\omega _{\alpha \beta }t}\left\langle \alpha ,\vec{p}%
\right\vert \vec{Q}_{S}\left\vert \beta ,\vec{p}\right\rangle \times
\left\langle \beta ,\vec{p}\right\vert \vec{v}\left\vert \alpha ,\vec{p}%
\right\rangle \text{.}
\end{eqnarray*}%
From Eq. (\ref{qonp}), we obtain%
\begin{eqnarray*}
\left\langle M\right\rangle &=&-i\frac{e}{2}\sum\limits_{\alpha ,p;\beta
,p}f\left( E_{\alpha }\left( \vec{p}\right) \right) e^{i\omega _{\alpha
\beta }t}\left\langle \alpha ,\vec{p}\right\vert \nabla _{\vec{k}}\left\vert
\beta ,\vec{p}\right\rangle \times \left\langle \beta ,\vec{p}\right\vert 
\vec{v}\left\vert \alpha ,\vec{p}\right\rangle , \\
&=&-i\frac{e}{2}\sum\limits_{\alpha ,p;\beta ,p}f\left( E_{\alpha }\left( 
\vec{p}\right) \right) e^{i\omega _{\alpha \beta }t}\left\langle \alpha ,%
\vec{p}\right\vert \nabla _{\vec{k}}\left\vert \beta ,\vec{p}\right\rangle
\times \left\langle \beta ,\vec{p}\right\vert \nabla _{\vec{k}}\left\vert
\alpha ,\vec{p}\right\rangle \omega _{\beta \alpha }\text{.}
\end{eqnarray*}%
Integrating the RHS with respect to time, the result is in units of orbital
magnetic moment multiplied by time denoted by $\left\langle \mathcal{\vec{M}}%
\right\rangle $, 
\begin{eqnarray*}
\left\langle \mathcal{\vec{M}}\right\rangle &=&-i\frac{e}{2}%
\sum\limits_{\alpha ,p;\beta ,p}f\left( E_{\alpha }\left( \vec{p}\right)
\right) \int\limits_{-\infty }^{0}e^{i\omega _{\alpha \beta }t}dt\ \omega
_{\beta \alpha }\left\langle \alpha ,\vec{p}\right\vert \nabla _{\vec{k}%
}\left\vert \beta ,\vec{p}\right\rangle \times \left\langle \beta ,\vec{p}%
\right\vert \nabla _{\vec{k}}\left\vert \alpha ,\vec{p}\right\rangle , \\
&=&-i\frac{e}{2}\sum\limits_{\alpha ,p;\beta ,p}f\left( E_{\alpha }\left( 
\vec{p}\right) \right) \frac{\omega _{\beta \alpha }}{i\omega _{\alpha \beta
}}dt\ \left\langle \alpha ,\vec{p}\right\vert \nabla _{\vec{k}}\left\vert
\beta ,\vec{p}\right\rangle \times \left\langle \beta ,\vec{p}\right\vert
\nabla _{\vec{k}}\left\vert \alpha ,\vec{p}\right\rangle , \\
&=&\frac{e}{2}\sum\limits_{\alpha ,p;\beta ,p}f\left( E_{\alpha }\left( \vec{%
p}\right) \right) \ \left\langle \alpha ,\vec{p}\right\vert \nabla _{\vec{k}%
}\left\vert \beta ,\vec{p}\right\rangle \times \left\langle \beta ,\vec{p}%
\right\vert \nabla _{\vec{k}}\left\vert \alpha ,\vec{p}\right\rangle , \\
&=&-\frac{e}{2}\sum\limits_{\alpha ,p;\beta ,p}f\left( E_{\alpha }\left( 
\vec{p}\right) \right) \ \left\langle \alpha ,\nabla _{\vec{k}}\vec{p}%
\right\vert \left\vert \beta ,\vec{p}\right\rangle \times \left\langle \beta
,\vec{p}\right\vert \nabla _{\vec{k}}\left\vert \alpha ,\vec{p}\right\rangle
, \\
&=&-\frac{e}{2}\sum\limits_{\alpha ,p}f\left( E_{\alpha }\left( \vec{p}%
\right) \right) \ \left\langle \alpha ,\nabla _{\vec{k}}\vec{p}\right\vert
\times \left\vert \alpha ,\nabla _{\vec{k}}\vec{p}\right\rangle \text{.}
\end{eqnarray*}%
The last line is just%
\begin{eqnarray*}
&&\left\langle \mathcal{\vec{M}}\right\rangle \\
&=&-\frac{e}{2}\frac{a^{2}}{\left( 2\pi \right) ^{2}}\sum\limits_{\alpha
}\int \int dk_{x}dk_{y}\ f\left( E_{\alpha }\left( \vec{k}\right) \right) %
\left[ 
\begin{array}{c}
\left\langle \frac{\partial }{\partial k_{x}}\alpha ,\vec{k}\right\vert 
\frac{\partial }{\partial k_{y}}\left\vert \alpha ,\vec{k}\right\rangle \\ 
-\left\langle \frac{\partial }{\partial k_{y}}\alpha ,\vec{k}\right\vert 
\frac{\partial }{\partial k_{x}}\left\vert \alpha ,\vec{k}\right\rangle%
\end{array}%
\right] , \\
&=&-\frac{e}{2}\frac{a^{2}}{\left( 2\pi \right) ^{2}}\sum\limits_{\alpha
}\int \int dk_{x}dk_{y}\ f\left( E_{\alpha }\left( \vec{k}\right) \right) %
\left[ 
\begin{array}{c}
\left\langle \frac{\partial }{\partial k_{x}}\alpha ,\vec{k}\right\vert 
\frac{\partial }{\partial k_{y}}\left\vert \alpha ,\vec{k}\right\rangle \\ 
-\left\langle \frac{\partial }{\partial k_{y}}\alpha ,\vec{k}\right\vert 
\frac{\partial }{\partial k_{x}}\left\vert \alpha ,\vec{k}\right\rangle%
\end{array}%
\right] , \\
&=&-\frac{e}{2}\frac{a^{2}}{\left( 2\pi \right) ^{2}}\sum\limits_{\alpha }\
f\left( E_{\alpha }\left( \vec{k}\right) \right) \int \int dk_{x}dk_{y}\
\left( \nabla _{\vec{k}}\times \left\langle \alpha ,\vec{k}\right\vert \frac{%
\partial }{\partial \vec{k}}\left\vert \alpha ,\vec{k}\right\rangle \right) 
\text{,}
\end{eqnarray*}%
explicitly revealing the Berry curvature derived from the expression for the
orbital magnetic moment.

\section{Concluding Remarks}

In summary, we have identified topological invariant in the effective $%
\left( \vec{p},\vec{q},E,t\right) -$phase space quantum transport,
explicitly given by 
\begin{eqnarray}
&&\frac{1}{\left( 2\pi \hbar \right) }\sum\limits_{occupied\text{ }%
bands}\int \int \int dk_{x}dk_{y}dt\left[ \frac{\partial ^{\left( a\right) }%
}{\partial k_{x}}\frac{\partial ^{\left( b\right) }}{\partial k_{y}}-\frac{%
\partial ^{\left( a\right) }}{\partial k_{y}}\frac{\partial ^{\left(
b\right) }}{\partial k_{x}}\right]  \nonumber \\
&&\times H_{occupied\text{ }band}^{\left( a\right) }\left( \mathcal{\vec{K}},%
\mathcal{E}\right) G^{<\left( b\right) }\left( \mathcal{\vec{K}},\mathcal{E}%
\right) ,  \label{phaseChern}
\end{eqnarray}%
an integral which give results in $%
\mathbb{Z}
$ manifold, the so-called first Chern numbers. This is explicitly
demonstrated here by reverting to its equivalent matrix elements expression.
Moreover, the conventional linearity in the electrical field strength 
\textit{per se} is not a necessary and sufficient condition to prove the
IQHE, but rather it is the first-order gradient expansion in the real-time
SFLWT-NEGF quantum transport equation.

In general, nonlinearity in the electric field is still present in the
variable $\left( \mathcal{\vec{K}},\mathcal{E}\right) $ in the integrand of
Eq. (\ref{hallcond}) based on the assumption of weak electric field and
hence weak time dependence. This aspect of our theory cannot arise from
linear response theory of equilibrium system. This is one of the virtues of
nonequilibrium quantum transport approach. There seems to be experimental
evidence on this nonlinearity \cite{Schade} of quantum Hall effect.

The fundamental physics underlying IQHE is the Berry connection and Berry
curvature, which is related to the geometrical origin of the old
Bohr-Sommerfeld quantization \cite{fego}. There have been other attempts to
formulate topological invariants and IQHE \cite{zubkov, shitade} in phase
space, but the results were not explicitly reduced to an expression in terms
of the physics of Berry connection, curvature and chern number, i.e., within
the fundamental quantization procedure first enunciated by the
Bohr-Sommerfeld quantization \cite{fego}, in a manner similar to was done in
reducing Eq. (\ref{phaseChern}) to Eq. (\ref{bingo}), or to their explicit
geometrical equivalence, such as, e.g., eigensheaves of the Hecke operators
in geometric Langlands program given by Ikeda \cite{ikeda}. Moreover, the
incorporation of varying potential and varying magnetic fields \cite{zubkov}
needs self-consistency with Maxwell's equations, which can only be attained
through large-scale and iterative numerical simulations incorporating
nonlinear transport feedback \cite{bj}, when IQHE is viewed as a \textit{%
bona fide} transport problem.

Therefore, we cannot compare our results with the published expressions of
these authors \cite{zubkov, shitade} for a couple of specific reasons, (a)
there is no transparency given towards Berry connection and Berry curvature
as indicated above, (b) Weyl transformation is as old as quantum mechanics,
and other phase space formulation are based on the old Moyal expansion. The
Buot formalism of this paper is physically founded on eigenfunction of
position and momentum operators. The discrete phase space is rigorously
based on the modular arithmetic of finite field \cite{gibbons}, a
generalization of the Born-von Karman boundary condition of solid-state
physics. It generalizes Wigner's continuum phase space formulation of
quantum mechanics by adapting to the problem at hand, specifically in terms
of the eigenfunctions for the lattice position and crystal momentum, namely
Wannier function (conventional, electric and magnetic) and corresponding
Bloch functions (conventional, electric and magnetic) leading to a
renormalized discrete phase space on finite fields \cite{gibbons, buotbook}.

It is worth mentioning that there are other more recent formulations of
Weyl-Wigner formulation of lattice models in quantum mechanics \cite%
{fialkovs, kasper, liga}. It should be mentioned at the outset that these
formulations are not explicitly based on modular arithmetic of finite field
modulo prime number of discrete lattice points ( i.e., Born-von Karman
boundary condition of solid-state physics with closure properties), which
could potentially create some ambiguities between discrete lattice and the
adopted continuous momentum variables, a bijective deficiency \cite{comments}%
. The Buot's formulation of discrete lattice points in a crystalline solid
obeying the Born-von Karman boundary condition is fundamentally based on a
finite field represented by a finite prime modulus number of lattice points,
in all directions on a 'torus'. All manipulations obey modular arithmetic,
closed under addition, multiplication, and division, where division by all
nonzero numbers is possible if and only if the modulus is a prime number. In
the terminology of abstract algebra, the ability to perform division means
that modular arithmetic modulo a prime number forms a field or, more
specifically, a finite field. This aspect is inherent in all the arithmetic
maipulations of the Buot formalism, and have been put to a firm theoretical
foundation by Gibbons et al \cite{gibbons}. This closure property for a
finite number of discrete points, only afforded by the modular arithmetic of
finite fields \cite{comments} that settled the theoretical aspects of
discrete quantum mechanics, is missing in all the recent papers on
Weyl-Wigner formulation of lattice models \cite{fialkovs, kasper, liga}. A
more complete discussions on various aspects of Buot's discrete lattice
formulations \footnote{%
Interested readers are referred to the following chapters of the author's
book:
\par
[{\small Chapter 37: Operator Hilbert-Space Methodology in Quantum Physics }
\par
{\small Chapter 38: The Wigner Function Construction}
\par
{\small Chapter 39: Discrete Phase Space on Finite Fields Chapter 40:
Discrete Quantum Mechanics on Finite Fields; and}
\par
{\small Chapter 41: Discrete Wigner Function Construction]}} is covered in
more details in the author's book \cite{buotbook} and references therein%
{\small .}

The key to the straightforward simplicity and power unique to Buot formalism
lies in the use of appropriate basis states for lattice position and crystal
momentum that make up the phase space, akin to a \textit{renormalization of}
phase space. It can even deal with a lattice system of only two discrete
points yielding Pauli matrices and the well-known Hadamard transformation
for qubits. In contrast, the method of Fialkovsky et al \cite{fialkovs}
failed to treat the simplest two-site systems \cite{comments}.

In the presence of uniform magnetic field, basically the canonical momentum
in Eq. (\ref{currenteq}) will strictly incorporate the magnetic field,
through the vector potential \cite{bj} in $\mathcal{\vec{K}}$. Thus, the
formalism immediately transfers to that of free electron gas under intense
magnetic fields, upon which the beautiful pioneering experiments of von
Klitzing, Dorda, and Pepper \cite{kdp} was performed. This
interchangeability is one of the advantages of our approach. It also appears
that electron-electron interaction \cite{bj} which does not break the
symmetry of Eqs. (\ref{eq2})-(\ref{eq3}) can be treated in similar manner.
The real-time SFLWT-NEGF multi-spinor quantum transport equations are also
able to predict various entanglements leading to different topological
phases of low-dimensional and nanostructured gapped condensed matter systems 
\cite{physicab}.

The method employed\footnote{%
For the unfamiliar readers of the new formalism, we give the following guide.
\par
For the theoretical formalism, the readers are directed to Refs. \cite{trHn}%
, \cite{fab}-\cite{fab6}, \cite{wannier} and for aspects of the formalism as
explicit discrete quantum theory on finite fields, Refs.\cite{trHn}, \cite%
{gibbons} and particularly chapter $39$ of Buot's book \cite{buotbook}
should be consulted.
\par
For some of the various applications, the readers are referred to Refs. \cite%
{bj, bok, fego, physicab, fabis, jb, rossi}.} is based on Buot's SFLWT-NEGF
formalism \cite{buotbook}. The use of the proper basis states is the key to
the simplicity of the physically based Buot discrete phase space formalism 
\cite{trHn} compared to others \cite{comments}. Our new approach to IQHE is
a \textit{nonequilibrium quantum transport based}, in contrast with
linear-response theory of equilibrium system employed so far. It seems more
appropriate for treating nonequilibrium transport of the IQHE. It
incorporates residual nonlinearity in $\mathcal{\vec{K}}$ and bypasses the
use of linear response theory and Kubo formula for the current-current
correlation which is based on time-dependent perturbation of equilibrium
system, thus the need to take the $\omega $ goes to $0$ limit.

We believe that this is the first time that real-time SFLWT-NEGF quantum
transport formalism is demonstrated to yield the transport-based topological
invariants in phase space that is transparent to the real physics of
quantization \cite{fego}. In another publication \cite{physicab}, the
real-time SFLWT-NEGF multi-spinor quantum transport equations lead to
prediction of various entanglements of different topological phases of
low-dimensional and nanostructured gapped condensed matter systems.

\begin{acknowledgement}
The author is grateful for the support of the Balik Scientist Program of the
Philippine Council for Industry, Energy and Emerging Technology Research and
Development (PCIEERD), Department of Science and Technology (DOST) and for
the hospitality of the USC Department of Physics as a Visiting Professor.
\end{acknowledgement}

\appendix

\section{Renormalized Bloch-electron Hamiltonian}

In this appendix we employ bold characters for vector quantities. In the
following, let us consider the crystal-lattice effective Hamiltonian for
energy band $n$ in the absence of uniform electric field using LWT given by 
\cite{buotbook, trHn} 
\begin{equation}
\mathcal{H}_{renormalized}=E_{n}\left( \mathbf{\hat{P}}\right) -e\vec{F}%
\cdot \mathbf{\hat{Q},}  \nonumber
\end{equation}%
where we have used the zero-field or conventional Wannier function and Bloch
functions of energy band theory, to obtain the LWT. In Eq. (\ref{eqLWT}), $%
\mathbf{\hat{P}}$ is the crystal momentum operator and $\mathbf{\hat{Q}}$ is
the lattice position operator, with Bloch eigenfunction and Wannier
eigenfunction, respectively. $E_{n}\left( \mathbf{p}\right) $ is the energy
band function for the band index $n$. Observe that the effective
(renormalized) Hamiltonian is simply obtained by the replacement of \textit{%
bare} momentum operator by the \textit{crystal momentum} operator and 
\textit{bare} coordinate operator replaced by the \textit{lattice} position
operator, a sort of \textit{renormalization} of phase space obtained through
the lattice Weyl transformation technique \cite{trHn}.

In the presence of uniform electric fields, if we substitute the solution of
Eq. (\ref{discuss}) derived in Appendix B, we see that eigenvalue of $%
H_{renormalized}$ is a function of%
\[
\mathbf{K}=\mathbf{p}+e\mathbf{F}t 
\]
and%
\[
\mathcal{E=}E_{o}+e\mathbf{F}\cdot \mathbf{q}. 
\]
We denote the electric Bloch state vector which is an eigenvector of the
crystal \textit{canonical} momentum operator $\mathbf{\hat{K}}$ by $%
\left\vert \mathbf{K}\right\rangle $ and the electric Wannier state vector
which is an eigenfunction of the lattice position operator $\mathbf{\hat{Q}}$
by $\left\vert \mathbf{q}\right\rangle $. Of course, the electric Bloch
function is given by $\left\langle \mathbf{x}\right\vert \left. \mathbf{K}%
\right\rangle $ and the electric Wannier function is given by $\left\langle 
\mathbf{x}\right\vert \left. \mathbf{q}\right\rangle $. Note that the
electric Wannier function carries a 'Peierls phase factor' given by Eq. (\ref%
{phsefactorq}) below. We have, by suppressing the band indices, 
\[
\mathbf{\hat{Q}}\left\vert \mathbf{q}\right\rangle =\mathbf{q}\left\vert 
\mathbf{q}\right\rangle , 
\]%
\[
\mathbf{\hat{K}}\left\vert \mathbf{q}\right\rangle =i\hbar \nabla _{\mathbf{q%
}}\left\vert \mathbf{q}\right\rangle \text{,} 
\]%
\[
\mathbf{\hat{K}}\left\vert \mathbf{K}\right\rangle =\mathbf{K}\left\vert 
\mathbf{K}\right\rangle \text{,} 
\]%
\begin{equation}
\mathbf{\hat{Q}}\left\vert \mathbf{K}\right\rangle =-i\hbar \nabla _{\mathbf{%
\ K}}\left\vert \mathbf{K}\right\rangle \text{.}  \label{qonp}
\end{equation}%
Note that $\left\vert \mathbf{K}\right\rangle $ can be considered also as
eigenfunction of the crystal momentum operator, $\mathbf{\hat{P},}$ by
virtue of the combination, $\mathbf{K}=\mathbf{p}+e\mathbf{F}t$. Here, in
the presence of uniform electric field, the minimal coupling is dictated by
the electric Bloch function, eigenfunction of both the Hamiltonian and
electric translation operator, whose wave vector incorporates the so-called
minimal coupling, see Eq. (\ref{discuss}) and the discussion that follows.
The energy variable of the theory, $\mathcal{E}$ in Eq. (\ref{energyvar}),
which occur in time-dependent problems, is dictated by the nature of \textit{%
electric translation operator} in space and time, defined through the
spatio-temporal \textit{action principle}, Eq. (\ref{translationopST}).
Thus, really the overall eigenfunction \cite{bj}, $W_{n}\left( \mathbf{K,}%
\mathcal{E}\right) ,$ is an explicit function of crystal canonical momentum $%
\mathbf{K}$ and $\mathcal{E}$ discussed in more detail in Appendix $B$. In
general, the LWT of quantities nonlocal in space and time, i.e., $\left( 
\mathbf{x},\mathbf{x}^{\prime };t,t^{\prime }\right) $, is expected to
depend on all the variables $\left( \mathbf{p},\mathbf{q};E,t\right) $. We
shall see in Appendix $B$ that in our present problem, $\mathbf{p}$ and $t$
are combined in $\mathbf{K}$, whereas $E$ and $\mathbf{q}$ are combined in $%
\mathcal{E}$. The dependence on $\left( \mathbf{p},\mathbf{q};E,t\right) $
thus reduces to that of $\left( \mathbf{K},\mathcal{E}\right) $ as will be
shown below. Only in the case for inhomogeneous and non-steady state or
transient problems do the dependence of $\left( \mathbf{p},\mathbf{q}%
;E,t\right) $ becomes irreducible to $\left( \mathbf{K},\mathcal{E}\right) $%
. On the other hand, for pure translational symmetric and steady-state
systems, $\left( \mathbf{p},\omega \right) $ goes to$\left( \mathbf{K},%
\mathcal{E}\right) $ (in a uniform electric field), so that only $\left( 
\mathbf{K},\mathcal{E}\right) $ dependence occurs in the pertinent LWT.

\subsection{Lattice Weyl transformation}

It is helpful to first discuss the zero-electric field Bloch function and
Wannier function, which are related via discrete Fourier transformation 
\begin{eqnarray}
b_{\lambda }\left( \mathbf{x},\mathbf{p}\right) &=&\left( N\hbar ^{3}\right)
^{-\frac{1}{2}}\sum\limits_{\mathbf{q}}e^{\left( \frac{i}{\hbar }\right) 
\mathbf{p}\cdot \mathbf{q}}w_{\lambda }\left( \mathbf{x},\mathbf{q}\right) ,
\nonumber \\
w_{\lambda }\left( \mathbf{x},\mathbf{q}\right) &=&\left( N\hbar ^{3}\right)
^{-\frac{1}{2}}\sum\limits_{\mathbf{p}}e^{-\left( \frac{i}{\hbar }\right) 
\mathbf{p}\cdot \mathbf{q}}b_{\lambda }\left( \mathbf{x},\mathbf{p}\right) .
\label{fouriertrans}
\end{eqnarray}%
The lattice Weyl transform of any operator $\hat{A}$ is thus given by 
\[
A_{\lambda \lambda ^{\prime }}\left( \mathbf{p},\mathbf{q}\right)
=\sum\limits_{\mathbf{v}}e^{\left( \frac{2i}{\hbar }\right) \mathbf{p}\cdot 
\mathbf{v}}\left\langle \mathbf{q}-\mathbf{v,\lambda }\right\vert \hat{A}%
\left\vert \mathbf{q}+\mathbf{v,\lambda }^{\prime }\right\rangle , 
\]%
which is also given by 
\[
A_{\lambda \lambda ^{\prime }}\left( \mathbf{p},\mathbf{q}\right)
=\sum\limits_{\mathbf{u}}e^{\left( \frac{2i}{\hbar }\right) \mathbf{q}\cdot 
\mathbf{u}}\left\langle \mathbf{p}+\mathbf{u,\lambda }\right\vert \hat{A}%
\left\vert \mathbf{p}-\mathbf{u,\lambda }^{\prime }\right\rangle . 
\]

\begin{proof}
We can use Eq. (\ref{fouriertrans}) and write 
\begin{eqnarray*}
A_{\lambda \lambda ^{\prime }}\left( \mathbf{p},\mathbf{q}\right)
&=&\sum\limits_{\mathbf{u}}e^{\left( \frac{2i}{\hbar }\right) \mathbf{q}%
\cdot \mathbf{u}}\left\langle \mathbf{p}+\mathbf{u,\lambda }\right\vert \hat{%
A}\left\vert \mathbf{p}-\mathbf{u,\lambda }^{\prime }\right\rangle , \\
&=&\left( N\hbar ^{3}\right) ^{-1}\sum\limits_{\mathbf{u}}e^{\left( \frac{2i%
}{\hbar }\right) \mathbf{q}\cdot \mathbf{u}} \\
&&\times \sum\limits_{\mathbf{q}^{\prime \prime }}e^{-\left( \frac{i}{\hbar }%
\right) \left( \mathbf{p+u}\right) \cdot \mathbf{q}^{\prime \prime
}}\left\langle \mathbf{q}^{\prime \prime }\mathbf{\ ,\lambda }\right\vert 
\hat{A}\sum\limits_{\mathbf{q}^{\prime }}e^{\left( \frac{i}{\hbar }\right)
\left( \mathbf{p-u}\right) \cdot \mathbf{q}^{\prime }}\left\vert \mathbf{q}%
^{\prime }\mathbf{,\lambda }^{\prime }\right\rangle , \\
&=&\left( N\hbar ^{3}\right) ^{-1}\sum\limits_{\mathbf{u}}e^{\left( \frac{2i%
}{\hbar }\right) \mathbf{q}\cdot \mathbf{u}} \\
&&\times \sum\limits_{\mathbf{q}^{\prime \prime },\mathbf{q}^{\prime
}}e^{\left( \frac{i}{\hbar }\right) \mathbf{p}\cdot \left( \mathbf{q}%
^{\prime }-\mathbf{q}^{\prime \prime }\right) }e^{-\left( \frac{i}{\hbar }%
\right) \mathbf{u}\cdot \left( \mathbf{q}^{\prime }+\mathbf{q}^{\prime
\prime }\right) }\left\langle \mathbf{q}^{\prime \prime }\mathbf{,\lambda }%
\right\vert \hat{A}\left\vert \mathbf{q}^{\prime }\mathbf{,\lambda }^{\prime
}\right\rangle .
\end{eqnarray*}%
Now let 
\begin{eqnarray*}
\mathbf{q}^{\prime \prime } &=&\mathbf{q}-\mathbf{v}\text{, \ \ \ \ \ } \\
\mathbf{q}^{\prime } &=&\mathbf{q}+\mathbf{v,}
\end{eqnarray*}%
then 
\begin{eqnarray*}
\mathbf{q}^{\prime }-\mathbf{q}^{\prime \prime } &=&2\mathbf{v}\text{, \ \ \
\ } \\
\text{\ }\mathbf{q}^{\prime }+\mathbf{q}^{\prime \prime } &=&2\mathbf{q,}
\end{eqnarray*}%
and%
\begin{eqnarray*}
A_{\lambda \lambda ^{\prime }}\left( \mathbf{p},\mathbf{q}\right) &=&\left(
N\hbar ^{3}\right) ^{-1}\sum\limits_{\mathbf{u,v}}e^{\left( \frac{2i}{\hbar }%
\right) \mathbf{q}\cdot \mathbf{u}}e^{-\left( \frac{i}{\hbar }\right) 
\mathbf{u}\cdot 2\mathbf{q}}e^{\left( \frac{i}{\hbar }\right) \mathbf{p}%
\cdot 2\mathbf{v}}\left\langle \mathbf{q}-\mathbf{\ v,\lambda }\right\vert 
\hat{A}\left\vert \mathbf{q}+\mathbf{v,\lambda }^{\prime }\right\rangle , \\
&=&\left( N\hbar ^{3}\right) ^{-1}\sum\limits_{\mathbf{u}}e^{0}\sum\limits_{%
\mathbf{v}}e^{\left( \frac{i}{\hbar }\right) \mathbf{p}\cdot 2\mathbf{v}%
}\left\langle \mathbf{q}-\mathbf{v,\lambda }\right\vert \hat{A}\left\vert 
\mathbf{q}+\mathbf{v,\lambda }^{\prime }\right\rangle , \\
&=&\left( N\hbar ^{3}\right) ^{-1}\left( N\hbar ^{3}\right) \sum\limits_{%
\mathbf{v}}e^{\left( \frac{i}{\hbar }\right) \mathbf{p}\cdot 2\mathbf{v}%
}\left\langle \mathbf{q}-\mathbf{v,\lambda }\right\vert \hat{A}\left\vert 
\mathbf{q}+\mathbf{v,\lambda }^{\prime }\right\rangle , \\
&=&\sum\limits_{\mathbf{v}}e^{\left( \frac{i}{\hbar }\right) \mathbf{p}\cdot
2\mathbf{v}}\left\langle \mathbf{q}-\mathbf{v,\lambda }\right\vert \hat{A}%
\left\vert \mathbf{q}+\mathbf{v,\lambda }^{\prime }\right\rangle \text{. \ \
\ \ \ \ \ \ \ \ QED}
\end{eqnarray*}
\end{proof}

\begin{example}
(a) zero-field case: (i) For the lattice position operator, we have 
\begin{eqnarray*}
\mathbf{q}\left( \mathbf{p},\mathbf{q}\right) &=&\sum\limits_{\mathbf{v}%
}e^{\left( \frac{2i}{\hbar }\right) \mathbf{p}\cdot \mathbf{v}}\left\langle 
\mathbf{q}-\mathbf{v,\lambda }\right\vert \mathbf{\hat{Q}}\left\vert \mathbf{%
q}+\mathbf{v,\lambda }^{\prime }\right\rangle , \\
&=&\mathbf{q}\sum\limits_{\mathbf{v}}e^{\left( \frac{2i}{\hbar }\right) 
\mathbf{p}\cdot \mathbf{v}}\delta \left( \mathbf{v}\right) \equiv \mathbf{q.}
\end{eqnarray*}%
\qquad (ii) For the crystal momentum operator, 
\begin{eqnarray*}
\mathbf{p}^{2}\left( \mathbf{p},\mathbf{q}\right) &=&\sum\limits_{\mathbf{u}%
}e^{\left( \frac{2i}{\hbar }\right) \mathbf{q}\cdot \mathbf{u}}\left\langle 
\mathbf{p}+\mathbf{u,\lambda }\right\vert \mathbf{\hat{P}}^{2}\left\vert 
\mathbf{p}-\mathbf{u,\lambda }^{\prime }\right\rangle , \\
&=&\mathbf{p}^{2}\sum\limits_{\mathbf{u}}e^{\left( \frac{2i}{\hbar }\right) 
\mathbf{q}\cdot \mathbf{u}}\delta \left( \mathbf{u}\right) =\mathbf{p}^{2},
\end{eqnarray*}%
resulting in the replacement of lattice position operator and crystal
momentum operator by their eigenvalues, respectively. (iii) For identity
operator, 
\begin{eqnarray*}
\mathbf{I}\left( \mathbf{p},\mathbf{q}\right) &=&\sum\limits_{\mathbf{v}%
}e^{\left( \frac{2i}{\hbar }\right) \mathbf{p}\cdot \mathbf{v}}\left\langle 
\mathbf{q}-\mathbf{v,\lambda }\right\vert \mathbf{\hat{I}}\left\vert \mathbf{%
q}+\mathbf{v,\lambda }^{\prime }\right\rangle , \\
&=&\sum\limits_{\mathbf{v}}e^{\left( \frac{2i}{\hbar }\right) \mathbf{p}%
\cdot \mathbf{v}}\delta _{\lambda \lambda ^{\prime }}\left( v\right) =1\cdot
\delta _{\lambda \lambda ^{\prime }}, \\
&=&\sum\limits_{\mathbf{u}}e^{\left( \frac{2i}{\hbar }\right) \mathbf{q}%
\cdot \mathbf{u}}\left\langle \mathbf{p}+\mathbf{u,\lambda }\right\vert 
\mathbf{\hat{I}}\left\vert \mathbf{p}-\mathbf{u,\lambda }^{\prime
}\right\rangle , \\
&=&\sum\limits_{\mathbf{u}}e^{\left( \frac{2i}{\hbar }\right) \mathbf{q}%
\cdot \mathbf{u}}\delta _{\lambda \lambda ^{\prime }}\left( u\right) =1\cdot
\delta _{\lambda \lambda ^{\prime }}.
\end{eqnarray*}%
(b) Non-zero field case: Using the results given by Eq. (\ref{transfunc})
below, and incorporating the Peierls phase factor, the discrete Fourier
transformation becomes 
\begin{equation}
b_{\lambda }\left( \mathbf{x},\mathbf{p}\right) =\left( N\hbar ^{3}\right)
^{-\frac{1}{2}}\sum\limits_{\mathbf{q}}e^{-\left( \frac{i}{\hbar }\right) 
\mathbf{p}\cdot \mathbf{q}}\left\{ \exp \left[ -\frac{i}{\hbar }\left( e\vec{%
F}t\cdot \vec{q}\ \right) \right] w_{\lambda }\left( x,\mathbf{q}\right)
\right\} ,  \nonumber
\end{equation}%
becomes%
\[
b_{\lambda }\left( \mathbf{x},\mathbf{p+eF}t\right) =\left( N\hbar
^{3}\right) ^{-\frac{1}{2}}\sum\limits_{\mathbf{q}}e^{-\left( \frac{i}{\hbar 
}\right) \left( \mathbf{p+eF}t\right) \cdot \mathbf{q}}w_{\lambda }\left( x,%
\mathbf{q}\right) . 
\]%
The inverse transformation is thus 
\begin{eqnarray*}
\tilde{w}_{\lambda }\left( x,\mathbf{q}\right) &=&\left( N\hbar ^{3}\right)
^{-\frac{1}{2}}\sum\limits_{\mathbf{p}}e^{\left( \frac{i}{\hbar }\right) 
\mathbf{p}\cdot \mathbf{q}}b_{\lambda }\left( \mathbf{x},\mathbf{p+eF}%
t\right) , \\
&=&\exp \left[ -\frac{i}{\hbar }\left( e\mathbf{F}t\cdot \mathbf{q}\ \right) %
\right] \left\{ \left( N\hbar ^{3}\right) ^{-\frac{1}{2}}\sum\limits_{%
\mathbf{p}^{\prime }}e^{\left( \frac{i}{\hbar }\right) \mathbf{p}^{\prime
}\cdot \mathbf{q}}b_{\lambda }\left( \mathbf{x},\mathbf{p}^{\prime }\right)
\right\} ,
\end{eqnarray*}%
becomes%
\begin{eqnarray*}
&&\exp \left[ -\frac{i}{\hbar }\left( e\mathbf{F}t\cdot \mathbf{q}\ \right) %
\right] \ w_{\lambda }\left( x,\mathbf{q}\right) \\
&=&\left( N\hbar ^{3}\right) ^{-\frac{1}{2}}\sum\limits_{\mathbf{p}%
}e^{\left( \frac{i}{\hbar }\right) \mathbf{p}\cdot \mathbf{q}}\ b_{\lambda
}\left( \mathbf{x},\mathbf{p+eF}t\right) ,
\end{eqnarray*}%
where $e^{-\left( \frac{i}{\hbar }\right) \mathbf{p}\cdot \mathbf{q}}$ is
the transition function between eigenfunctions of lattice positon operator
and crystal momentum operators derived below.
\end{example}

\subsubsection{Transition function: discrete Fourier transform}

Here we derived the \textit{general} transition function between
eigenvectors of lattice position operator and crystal momentum operators, $%
\left\vert \mathbf{q}\right\rangle $ and $\left\vert \mathbf{p}\right\rangle 
$, respectively. We have the identity, 
\begin{eqnarray}
\left\vert \mathbf{p}\right\rangle &=&\sum\limits_{q}\left\vert \mathbf{q}%
\right\rangle \left\langle \mathbf{q}\right\vert \left\vert \mathbf{p}%
\right\rangle ,  \nonumber \\
&=&\sum\limits_{q}\left\langle \mathbf{q}\right\vert \left\vert \mathbf{p}%
\right\rangle \ \left\vert \mathbf{q}\right\rangle \text{.}  \label{trans1}
\end{eqnarray}%
Similarly 
\begin{eqnarray}
\left\vert \mathbf{q}\right\rangle &=&\sum\limits_{p}\left\vert \mathbf{p}%
\right\rangle \left\langle \mathbf{p}\right\vert \left\vert \mathbf{q}%
\right\rangle ,  \nonumber \\
&=&\sum\limits_{p}\left\langle \mathbf{p}\right\vert \left\vert \mathbf{q}%
\right\rangle \ \left\vert \mathbf{p}\right\rangle  \label{trans2}
\end{eqnarray}%
Eqs. (\ref{trans1}) and (\ref{trans2}) define discrete Fourier
transformation. To derive the functional form of $\left\langle \mathbf{p}%
\right\vert \left\vert \mathbf{q}\right\rangle $ or its complex conjugate $%
\left\langle \mathbf{q}\right\vert \left\vert \mathbf{p}\right\rangle $, we
make use of the following relations on wavevectors (\textit{not wavefunctions%
}): 
\[
\mathbf{\hat{Q}}\left\vert \mathbf{p}\right\rangle \Rightarrow -i\hbar
\nabla _{\mathbf{p}}\left\vert \mathbf{p}\right\rangle , 
\]%
and%
\[
\mathbf{\hat{P}}\left\vert \mathbf{q}\right\rangle =i\hbar \nabla _{\mathbf{q%
}}\left\vert \mathbf{q}\right\rangle . 
\]%
Therefore, it follows that 
\begin{eqnarray}
\left\langle \mathbf{q}\right\vert \mathbf{\hat{Q}}^{\dagger }\left\vert 
\mathbf{p}\right\rangle &=&\mathbf{q}\left\langle \mathbf{q}\right\vert
\left\vert \mathbf{p}\right\rangle ,  \nonumber \\
\left\langle \mathbf{q}\right\vert -i\hbar \nabla _{\mathbf{p}}\left\vert 
\mathbf{p}\right\rangle &=&\mathbf{q}\left\langle \mathbf{q}\right\vert
\left\vert \mathbf{p}\right\rangle ,  \nonumber \\
-i\hbar \nabla _{\mathbf{p}}\left\langle \mathbf{q}\right\vert \left\vert 
\mathbf{p}\right\rangle &=&\mathbf{q}\left\langle \mathbf{q}\right\vert
\left\vert \mathbf{p}\right\rangle ,  \nonumber \\
\nabla _{\mathbf{p}}\left\langle \mathbf{q}\right\vert \left\vert \mathbf{p}%
\right\rangle &=&\left( \frac{i}{\hbar }\mathbf{q}\right) \left\langle 
\mathbf{q}\right\vert \left\vert \mathbf{p}\right\rangle ,  \nonumber \\
\ln \left\langle \mathbf{q}\right\vert \left\vert \mathbf{p}\right\rangle
&=&\left( \frac{i}{\hbar }\mathbf{p}\cdot \mathbf{q}\right) ,\text{ \ } 
\nonumber \\
\text{\ }\left\langle \mathbf{q}\right\vert \left\vert \mathbf{p}%
\right\rangle &=&const\cdot \exp \left( \frac{i}{\hbar }\mathbf{p}\cdot 
\mathbf{q}\right) ,  \nonumber \\
\left[ const\cdot \exp \left( \frac{i}{\hbar }\mathbf{p}\cdot \mathbf{q}%
\right) \right] ^{\ast } &=&\left\langle \mathbf{p}\right\vert \left\vert 
\mathbf{q}\right\rangle =const\cdot \exp \left( -\frac{i}{\hbar }\mathbf{p}%
\cdot \mathbf{q}\right) .  \label{transfunc}
\end{eqnarray}%
where 
\[
const=\left( N\hbar ^{3}\right) ^{-\frac{1}{2}}. 
\]%
The use of partial differentiation in Eq. (\ref{transfunc}) is justified
since in the limit of infinite volume $\mathbf{p}$ becomes continuous. The
use of partial differentiation in the case of lattice point $\mathbf{q}$
which is involved in deriving $\left\langle \mathbf{p}\right\vert \left\vert 
\mathbf{q}\right\rangle $ can be based on the assumption that there exist
continuous functions of $\mathbf{q}$ having infinite radius of convergence,
which are equal to the correct values at the lattice sites.

\section{Spatio-temporal translation operators (STTO): action phase operators%
}

Note that displacement operators are characteristically exponential
operators amenable to Fourier series expansion. Remarkably, phase factors
are generally acquired due to displacement or motion in parameter space in
the presence of electric or magnetic fields. This is ubiquitous in
solid-state physics (e.g., Peierls phase factor in magnetic fields) before
the Berry connection become mainstream and fashionable. To begin, we write
our bare Hamiltonian of electrons as 
\[
H=H_{o}-e\vec{F}\cdot \vec{r}, 
\]%
where $H_{o}$ is the periodic Hamiltonian in the absence of the electric
field, $\vec{F}$. In what follows, we will prove the following results: (a)
displacement operators in space and time carry their respective "Peierls"
phase factors, (b) the time derivative of momentum operator is equal to $e%
\vec{F}\cdot \vec{q}$, and (c) gradient of the energy operator equals $e\vec{%
F}t$. Specifically we want to show that the displacement operator, 
\[
\hat{T}\left( -q\right) =\exp \left[ -\frac{i}{\hbar }\left( \left( \vec{q}%
\cdot -i\hbar \nabla _{\vec{r}}\right) \right) \right] , 
\]%
acquires a phase factor in the presence of electric field, and is given by 
\begin{eqnarray}
\tilde{T}\left( -\vec{q}\right) &=&\exp \left[ -\frac{i}{\hbar }\left(
\left( e\vec{F}\cdot \vec{q}\right) t+\left( \vec{q}\cdot -i\hbar \nabla _{%
\vec{r}}\right) \right) \right] ,  \nonumber \\
&=&\exp \left[ -\frac{i}{\hbar }\left( \left( e\vec{F}\cdot \vec{q}\right)
t+\left( \vec{q}\cdot \hat{P}\right) \right) \right] ,  \label{spaceT}
\end{eqnarray}%
where we indicate by $\tilde{T}$ the translation operator $\hat{T}$ with a
phase factor.

Similarly, the time displacement operator, 
\[
\hat{T}\left( t\right) =\exp \left[ -\frac{i}{\hbar }\left( \left( t\cdot \
i\hbar \frac{\partial }{\partial t}\right) \right) \right] , 
\]
also acquires a phase factor and is given by 
\begin{equation}
\tilde{T}\left( t\right) =\exp \left[ -\frac{i}{\hbar }\left( \left( e\vec{F}%
\cdot \vec{q}\right) t+\left( t\cdot \ i\hbar \frac{\partial }{\partial t}%
\right) \right) \right] ,  \label{timeT}
\end{equation}%
where $\tilde{T}\left( t\right) $ differs from $\hat{T}\left( t\right) $ by
a phase factor, and $i\hbar \frac{\partial }{\partial t}$ is the energy
operator of the system. In other words, displacement in space by lattice
vector $-q$ acquires phase factors equal to $\exp \left( -\frac{i}{\hbar }e%
\vec{F}t\cdot \vec{q}\right) $. Smilarly, displacement in time by $t$
acquires phase factor equal to $\exp \left( -\frac{i}{\hbar }\left( e\vec{F}%
\cdot \vec{q}\right) t\right) $. In the absence of the electric field these
phase factors give unity, e.g., $\tilde{T}\left( t\right) \Longrightarrow _{%
\vec{F}\rightarrow 0}\hat{T}\left( t\right) $, reduces to ordinary
translation operator.

The physics behind these phase factors is dictated by a selfconsistent
translation of local functions in space and time under a uniform electric
field, $\vec{F}$. These phases are explained in more detail in the next
section. For efficient bookeeping and for taking lattice Weyl transform, it
is usually more convenient to attach these phase factors to displaced local
functions themselves. This yields a generalization of Peierls phase factor,
well-known for solid-state problems in magnetic fields. The derivation goes
as follows.

\subsection{Derivation of the STTO based on action principle}

In Hamiltonian-Lagrangian mechanics, $\mathcal{L}=K.E.-V$, where $K.E.$ is
the kinetic energy and 
\begin{equation}
H=\sum\limits_{q_{i}}\dot{q}_{i}\frac{\partial \mathcal{L}}{\partial \dot{q}%
_{i}}-\mathcal{L}.  \label{hamilton}
\end{equation}%
The action is given by 
\begin{equation}
\int dt\mathcal{L}=\int dt\left( \sum\limits_{q_{i}}p_{i}\dot{q}%
_{i}-H\right) ,  \label{action}
\end{equation}%
which gives the Euler-Lagrange classical equation of motion given by the
stationary action principle, 
\begin{equation}
\frac{d}{dt}\frac{\partial \mathcal{L}}{\partial \dot{q}_{i}}-\frac{\partial 
\mathcal{L}}{\partial q_{i}}=0,  \label{EuLa}
\end{equation}%
which is the classical Newton force equation.

For our single particle case, a translation in space and time must have
inherent particle dynamics. The classical action of the displacement in
space and time then becomes operator as, 
\[
p\cdot q-Ht\Longrightarrow \hat{P}\cdot q-\hat{\xi}t 
\]%
Hence, we define translation operator in space and time, which commute i.e., 
$\left[ T\left( q\right) ,T\left( t\right) \right] =0$, as 
\begin{equation}
\hat{T}\left( q\right) \hat{T}\left( t\right) =\exp \frac{i}{\hbar }\left( 
\hat{P}\cdot q\right) \exp \frac{i}{\hbar }\left( -\hat{\xi}t\right) =\exp 
\frac{i}{\hbar }\left( \hat{P}\cdot q-\hat{\xi}t\right)
\label{translationopST}
\end{equation}%
where $\hat{P}$ is the momentum operator, 
\[
-i\hbar \nabla _{\mathbf{q}}\psi =\hat{P}\psi , 
\]%
and $\xi $ is the energy operator, explicitly given by 
\[
\hat{\xi}=i\hbar \frac{\partial }{\partial t} 
\]%
since in the Schr\"{o}dinger equation, 
\[
i\hbar \frac{\partial }{\partial t}\psi =\hat{H}\psi . 
\]%
In Eq. (\ref{translationopST}), the Baker--Campbell--Hausdorff formula has
been applied, since $\hat{T}\left( \vec{q}\right) $ and $\hat{T}\left(
t\right) $ commute and therefore equal to the right side of the equation.
The phase of the spatio-temporal translation operator mirrors that of the
Lagrangian, $\mathcal{L}$, in Hamilton classical mechanics, Eq. (\ref{action}%
), where the stationary action gives the classical equation of motion, Eq. (%
\ref{EuLa}). In quantum mechanics, this stationary action translates into
stationary quantum phase. With respect to our translation operator, we have
indeed a classical (i.e., using eigenvalues) equation of motion which gives
the classical distance (position) $q$ given by the group velocity multiplied
by the time traveled, $\vec{q}=\vec{v}_{g}t$ where%
\[
\vec{v}_{g}=\frac{\partial \mathcal{E}}{\partial \vec{p}} 
\]%
is the group velocity. Moreover, if we denote 
\[
L=\int\limits_{0}^{t}\left( p\cdot \dot{q}-\mathcal{E}\right) dt, 
\]%
then we have a stationary action principle, $\mathbf{\dot{p}}=e\mathbf{F}$
which describes the the impulse imparted on the particle. Thus, our proper
choice of translation operators in space and time, Eq. (\ref{translationopST}%
) has inherent particle dynamics built into it. Moreover, the phase of the
Schr\"{o}dinger wavefunction is also proportional to $L$, for describing a
traveling wavepacket, again justifying our spatio-temporal translation
operator, Eq. ( \ref{translationopST}).

\subsection{Peierls phase factors}

\subsubsection{Nonlocality in coordinates}

The nature of the derivatives of exponential displacement operator is
determined, e.g., by the following operation, 
\begin{eqnarray*}
i\hbar \frac{\partial }{\partial t}\tilde{T}\left( q\right) W_{\lambda
}\left( r-0\right) &=&i\hbar \frac{\partial \phi }{\partial t}\tilde{T}%
\left( q\right) W_{\lambda }\left( r-0\right) , \\
\frac{\partial }{\partial t}\tilde{T}\left( q\right) &=&\frac{\partial \phi 
}{\partial t}\tilde{T}\left( q\right) , \\
\frac{d\tilde{T}\left( q\right) }{\tilde{T}\left( q\right) } &=&d\phi ,
\end{eqnarray*}%
where specifically $\hat{T}\left( -q\right) W_{\lambda }\left( r-0\right) $
is a translation of the center coordinate of a localized Wannier function
centered in the origin to another lattice point $q$, yielding $W_{\lambda
}\left( r-q\right) .$ Therefore $\hat{T}\left( -q\right) $ resembles the
physical process of transfering a localized function centered at the origin
to a localized function centered at another lattice point $q$. Here we
define $\phi $ as undetermined for the moment as 
\[
\phi =-\frac{i}{\hbar }\left( f\left( q,t\right) +\left( \vec{q}\cdot \hat{P}%
\right) \right) , 
\]%
where $f\left( q,t\right) $ is to be determined. Note that the presence of $%
f\left( q,t\right) $ is needed for selfconsistency in the presence of
electric field. Now consider the second term in the exponent, namely, 
\begin{eqnarray*}
\phi _{2} &=&-\frac{i}{\hbar }\vec{q}\cdot \hat{P}=-\frac{i}{\hbar }\vec{q}%
\cdot \left( -i\hbar \frac{\partial }{\partial \vec{r}}\right) , \\
&=&\left( -q\right) \cdot \frac{\partial }{\partial \vec{r}}
\end{eqnarray*}%
leading to Fourier series expansion of the translation, $\hat{T}\left(
-q\right) $ . Then, we obtain 
\begin{equation}
-i\hbar \frac{\partial \phi _{2}}{\partial t}=\left[ H,\phi _{2}\right] =%
\frac{i}{\hbar }\left[ H,\left( \left( -\vec{q}\right) \cdot \hat{P}\right) %
\right] =e\vec{F}\cdot \left( -\vec{q}\right) ,  \label{eFq}
\end{equation}%
since $-\frac{i}{\hbar }f\left( q,t\right) $ commutes with the Hamiltonian.
Therefore, we have 
\begin{equation}
\frac{d\tilde{T}\left( -q\right) }{\tilde{T}\left( -q\right) }=d\phi =\frac{i%
}{\hbar }e\vec{F}\cdot \left( -\vec{q}\right) \ dt.  \label{devlog}
\end{equation}%
Therefore 
\begin{eqnarray}
\tilde{T}\left( -q\right) &=&\exp \frac{i}{\hbar }\left[ \left( e\vec{F}%
\cdot \left( -\vec{q}\right) \ \Delta t\right) +\hat{P}\cdot \left( -\vec{q}%
\right) \right] ,  \nonumber \\
&=&\exp \frac{i}{\hbar }\left[ \hat{P}+e\vec{F}t\right] \cdot \left( -\vec{q}%
\right)  \label{spacetransop}
\end{eqnarray}%
This means that a displacement $\left( -\vec{q}\right) $ of localized
function acquires a phase factor given by 
\begin{equation}
Peierls\ phase\ factor=\exp \left[ \frac{i}{\hbar }\left( e\vec{F}t\cdot
\left( -\vec{q}\right) \ \right) \right] .  \label{phsefactorq}
\end{equation}%
We may thus write 
\begin{equation}
-i\hbar \frac{\partial }{\partial t}\tilde{T}\left( -\vec{q},t\right) =\left[
H,\tilde{T}\left( -\vec{q},t\right) \right] =e\vec{F}\cdot \left( -\vec{q}%
\right) \ \tilde{T}\left( -\vec{q},t\right) .  \label{eq1}
\end{equation}%
We can also deduce from Eq. (\ref{eFq}) the relation for the momentum
operator, 
\begin{eqnarray}
\frac{i}{\hbar }\left[ H,\left[ \left( -\vec{q}\right) \cdot -i\hbar \nabla
_{\vec{r}}\right] \right] &=&e\vec{F}\cdot \left( -\vec{q}\right) =\frac{%
\partial \hat{P}}{\partial t}\cdot \left( -q\right) ,\Longrightarrow \frac{%
\partial \hat{P}}{\partial t}=e\vec{F},  \nonumber \\
&\Longrightarrow &\hat{P}=\hat{P}_{o}+e\vec{F}t.  \label{canonicalP}
\end{eqnarray}%
\bigskip

\subsubsection{Simultaneous eigenvalues for $\hat{H}$ and $\tilde{T}\left( -%
\vec{q},t\right) $}

One very important conclusion is implied in Eq. (\ref{eq1}), which we
rewrite here for convenience 
\begin{equation}
\left[ \hat{H},\tilde{T}\left( -\vec{q},t\right) \right] =-e\vec{F}\cdot 
\vec{q}\ \tilde{T}\left( -\vec{q},t\right) .  \label{discuss}
\end{equation}%
What the above relation means is that if $\tilde{T}\left( -\vec{q},t\right) $
is diagonal then $\left[ H,\tilde{T}\left( -\vec{q},t\right) \right] $ is
also diagonal. But if $\tilde{T}\left( -\vec{q},t\right) $ is diagonal, then 
$\hat{H}$ is also diagonal with the same eigenvalues. The eigenfunction of $%
\tilde{T}\left( -\vec{q},t\right) $ is labeled by the quantum label $%
\mathcal{\vec{K}}=\vec{p}_{o}+e\vec{F}t$. \ This implies that $\hat{H}$ is
also diagonal in $\mathcal{\vec{K}}$. The electric Bloch function labeled by 
$B\left( k_{o}+\frac{e}{\hbar }Ft,..\right) $ is the eigenfunction of $%
\tilde{T}\left( -\vec{q},t\right) $ as well as that of the renormalized 
\[
H_{renormalized}\left( \hat{K},\hat{Q}\right) \Longleftrightarrow
E_{n}\left( \mathcal{\vec{K}}\right) -e\vec{F}\cdot \vec{q}\ , 
\]
where the double pointed arrow denotes lattice Weyl correspondence, Eq. (\ref%
{eqLWT}), using the localized electric Wannier function.

\subsubsection{Nonlocality in time}

Because nonlocal arguments in time acquires phase factor also, the energy
variable of the theory now varies with $e\vec{F}\cdot \vec{q},$ as discussed
next. From Eq. (\ref{translationopST}) for a displacement in time 
\[
\hat{T}\left( t\right) =\exp \frac{i}{\hbar }\left( -\hat{\xi}t\right) 
\]%
or 
\[
\tilde{T}(t)\equiv \mathcal{F}\exp \left( \frac{-i}{\hbar }\left( i\hbar t%
\frac{\partial }{\partial t}\right) \right) =\mathcal{F}\exp \left( t\frac{%
\partial }{\partial t}\right) 
\]%
where, 
\begin{equation}
\mathcal{F}=\exp \left[ -\frac{i}{\hbar }\left( f_{t}\left( q,t\right)
\right) \right] ,  \label{timephase}
\end{equation}%
is a phase factor to be determined 
\[
\phi _{t}\left( q\right) =-\frac{i}{\hbar }\left( f_{t}\left( q,t\right) +t\
i\hbar \frac{\partial }{\partial t^{\prime }}\right) . 
\]%
Therefore,%
\begin{eqnarray}
-i\hbar \frac{\partial \tilde{T}(t)}{\partial \vec{q}} &=&-i\hbar \frac{%
\partial \phi _{t}\left( q\right) }{\partial \vec{q}}\tilde{T}(t),  \nonumber
\\
-i\hbar \frac{\partial \phi _{t}\left( q\right) }{\partial \vec{q}} &=&-%
\frac{i}{\hbar }\left[ \mathcal{\vec{K}}\left( t^{\prime }\right) ,\left(
t\right) i\hbar \frac{\partial }{\partial t^{\prime }}\right] =\left[ \vec{p}%
+\frac{e}{c}\vec{A}_{o},t\nabla _{t^{\prime }}\right] ,  \nonumber \\
&=&\left[ \vec{p}+e\vec{F}t^{\prime },t\nabla _{t^{\prime }}\right] =-e\vec{F%
}t.  \label{timedispla}
\end{eqnarray}%
Therefore 
\begin{eqnarray*}
-i\hbar \frac{\partial \phi _{t}\left( q\right) }{\partial \vec{q}} &=&-e%
\vec{F}t, \\
\frac{\partial \phi _{t}\left( q\right) }{\partial \vec{q}} &=&-\frac{i}{%
\hbar }e\vec{F}t.
\end{eqnarray*}%
Thus, we obtained, 
\[
\frac{\partial \tilde{T}(t)}{\partial \vec{q}}=\left( -\frac{i}{\hbar }e\vec{%
F}t\right) \tilde{T}(t), 
\]%
and hence,%
\begin{eqnarray*}
d\ln \tilde{T}(t) &=&\left( -\frac{i}{\hbar }e\vec{F}t\right) \cdot d\vec{q},
\\
\ln \tilde{T}(t) &=&\left( -\frac{i}{\hbar }e\vec{F}t\right) \cdot \Delta 
\vec{q}.
\end{eqnarray*}%
Hence a displacement in time carries a phase factor given by $\exp \left( -%
\frac{i}{\hbar }\left( e\vec{F}t\right) \cdot \Delta \vec{q}\right) $ and 
\begin{equation}
\tilde{T}(t)=\exp \left\{ -\frac{i}{\hbar }\left[ \left( e\vec{F}t\right)
\cdot \Delta \vec{q}+t\left( i\hbar \frac{\partial }{\partial t^{\prime }}%
\right) \right] \right\}  \label{timetransop}
\end{equation}%
Therefore, 
\begin{equation}
-i\hbar \frac{\partial }{\partial \vec{q}}\tilde{T}\left( \vec{q},t\right) =%
\left[ \mathcal{\vec{K}}\left( t\right) ,\tilde{T}\left( \vec{q},t\right) %
\right] =\left( -e\vec{F}t\right) \ \tilde{T}\left( \vec{q},t\right)
\label{eq2}
\end{equation}%
Now from second line of Eq. (\ref{timedispla}), we have 
\begin{eqnarray*}
-\frac{i}{\hbar }\left[ \mathcal{\vec{K}}\left( t^{\prime }\right) ,\left(
t\right) i\hbar \frac{\partial }{\partial t^{\prime }}\right] &=&\left[ \vec{%
p}+e\vec{F}t^{\prime },\left( t\right) \frac{\partial }{\partial t^{\prime }}%
\right] =-e\vec{F}t, \\
\left[ \mathcal{\vec{K}}\left( t^{\prime }\right) ,\hat{\xi}\right]
&=&-i\hbar e\vec{F}=-i\hbar \frac{\partial \hat{\xi}}{\partial q},
\end{eqnarray*}%
which leads to to the expression for in Eq. (\ref{timephase}) for $f_{t}$ in
the phase factor as 
\[
\frac{\partial \hat{\xi}}{\partial \vec{q}}=e\vec{F}\Longrightarrow f_{t}=e%
\vec{F}\cdot \vec{q}. 
\]%
All these results, Eq. (\ref{spacetransop}) and Eq. (\ref{timetransop}),
lead to the time-dependent wave vector, 
\[
\hbar \vec{k}=\hbar \vec{k}_{o}+\hbar e\vec{F}t 
\]
and to the position-dependent energy, 
\begin{equation}
\mathcal{E}=E_{o}+e\vec{F}\cdot \vec{q},  \label{energyvar}
\end{equation}%
respectively. Again, for for taking lattice Weyl transform, it is more
convenient to attach these phase factor to the displaced local functions
themselves. This allows us to generalize the Peierls phase factor to space
and time displacements, originally well-known for solid-state problems in
magnetic fields.

\section{Hall current density and current through a discrete lattice}

The definition of the current density given by the first row of Eq. (\ref%
{eqKE}), reads 
\begin{equation}
\frac{a^{2}}{\left( 2\pi \hbar \right) ^{2}}\diint dK_{x}\ dK_{y}\left( 
\frac{e}{a^{2}}\frac{\partial \mathcal{E}}{\partial K_{y}}\right) \left(
-iG^{<}\left( \mathcal{\vec{K}},\mathcal{E}\right) \right)  \label{first row}
\end{equation}%
One observes that the above expression embodies the intuitive concept of a
current density $j\left( E\right) $ as a product of group velocity,
represented by $v_{g}=\frac{\partial \mathcal{E}}{\partial K_{y}}$
multiplied by areal density of particles, represented by 
\[
\rho =\frac{1}{a^{2}}\left( -iG^{<}\left( \mathcal{\vec{K}},\mathcal{E}%
\right) \right) , 
\]
multiplied by the electric charge, $e$, and sumed over all the
two-dimensional variables, $K_{x}$ and $K_{y}$ with proper counting of
states. It can easily be shown that Eq. (\ref{first row}) has a
corresponding discrete lattice expression. Indeed, we can readily use the
expression for the \textit{directional} electric current through the lattice
derived in detail in Sec. $21.3$ of Ref. \cite{buotbook}, namely, Eq. ($%
21.16 $) in Sec. $21.3$ of the book, which give the current through the
lattice as 
\begin{eqnarray}
j\left( E,t\right) &=&\frac{-e}{\hbar }\sum\limits_{\vec{q},\vec{q}^{\prime
\prime }}\left( \vec{q}-\vec{q}^{\prime \prime }\right) \left\langle \vec{q}%
\right\vert H_{o}\left\vert \vec{q}^{\prime \prime }\right\rangle
G^{<}\left( \vec{q}^{\prime \prime },\vec{q},E,t\right) ,  \nonumber \\
&=&\frac{e}{\hbar }\sum\limits_{q,q^{\prime \prime }}\left( \vec{q}^{\prime
\prime }-\vec{q}\right) \left\langle \vec{q}\right\vert H_{o}\left\vert \vec{%
q}^{\prime \prime }\right\rangle G^{<}\left( \vec{q}^{\prime \prime },\vec{q}%
,E,t\right) .  \label{currentdensity}
\end{eqnarray}%
The above equation involves summation over all the lattice points, $q$ and $%
q^{\prime \prime }$, i.e., the expression is really a trace. Now a trace
involving products of matrix elements is also a trace involving products of
their respective lattice Weyl transforms \cite{trHn}. In what follows we
will treat steady-state condition only. The lattice Weyl transform of 
\[
\left( \vec{q}^{\prime \prime }-\vec{q}\right) \left\langle \vec{q}%
\right\vert H_{o}\left\vert \vec{q}^{\prime \prime }\right\rangle 
\]
can be calculated as follows. We have, 
\begin{eqnarray*}
\mathcal{W}\left[ \left( \vec{q}^{\prime \prime }-\vec{q}\right)
\left\langle \vec{q}\right\vert H_{o}\left\vert \vec{q}^{\prime \prime
}\right\rangle \right] &=&\sum\limits_{\vec{v}}e^{\left( \frac{2i}{\hbar }%
\right) \vec{p}\cdot \vec{v}}\left( \left( \tilde{q}+\vec{v}\right) -\left( 
\tilde{q}-\vec{v}\right) \right) \left\langle \tilde{q}-\vec{v}\right\vert
H_{o}\left\vert \tilde{q}+\vec{v}\right\rangle , \\
&=&\sum\limits_{\vec{v}}e^{\left( \frac{2i}{\hbar }\right) \vec{p}\cdot \vec{%
v}}\left( 2\vec{v}\right) \left\langle \tilde{q}-\vec{v}\right\vert
H_{o}\left\vert \tilde{q}+\vec{v}\right\rangle , \\
&=&\frac{\partial }{\partial \vec{p}}\sum\limits_{\vec{v}}e^{\left( \frac{2i%
}{\hbar }\right) \vec{p}\cdot \vec{v}}\left\langle \tilde{q}-\vec{v}%
\right\vert H_{o}\left\vert \tilde{q}+\vec{v}\right\rangle .
\end{eqnarray*}%
However, for crystalline solid the matrix element of the energy band depends
in the difference of positions $\vec{q}$ which is equivalent to saying that
the matrix element $\left\langle \tilde{q}-\vec{v}\right\vert
H_{o}\left\vert \tilde{q}+\vec{v}\right\rangle \Longrightarrow H_{o}\left( 2%
\vec{v}\right) $. Therefore we have 
\[
\mathcal{W}\left( \vec{q}^{\prime \prime }-\vec{q}\right) \left\langle \vec{q%
}\right\vert H_{o}\left\vert \vec{q}^{\prime \prime }\right\rangle =\frac{%
\partial }{\partial \vec{p}}E\left( \vec{p}\right) . 
\]%
In the presence of uniform electric or magnetic fields, we then have,%
\[
\frac{\partial }{\partial \vec{p}}E\left( \vec{p}\right) \Longrightarrow 
\frac{\partial }{\partial \mathcal{\vec{K}}}\mathcal{E}\left( \mathcal{\vec{K%
}}\right) . 
\]
Similarly, the lattice Weyl transform of $G^{<}\left( \vec{q}^{\prime \prime
},\vec{q},E,t\right) $ is 
\[
W\left[ -iG^{<}\left( \vec{q}^{\prime \prime },\vec{q},E,t\right) \right]
=-iG^{<}\left( \vec{p},\vec{q},E,t\right) , 
\]%
which goes into 
\[
-iG^{<}\left( \vec{p},\vec{q},E,t\right) \Longrightarrow -iG^{<}\left( 
\mathcal{\vec{K}},\mathcal{E}\right) , 
\]%
in the presence of electric or magnetic field. Using the proper counting of
states in phase space for two-dimensional systems in the taking of the
trace, we finally arrive at 
\[
j_{y}=\frac{a^{2}}{\left( 2\pi \hbar \right) ^{2}}\diint dK_{x}dK_{y}\left( 
\frac{e}{a^{2}}\frac{\partial \mathcal{E}}{\partial K_{y}}\right) \left(
-iG^{<}\left( \mathcal{\vec{K}},\mathcal{E}\right) \right) , 
\]%
which is what we want to derive from the current density through the
lattice, Eq. (\ref{currentdensity}). Thus, this equation is valid for both
lattice and continuum by simply performing lattice Weyl transformation of
the appropriate quantities from discrete to continuum and vice versa.

\section{Phase change of wavefunction under parallel transport}

To relate to the phase of the wavefunction, we recall that for parallel
transport 
\begin{eqnarray*}
\frac{\partial \psi _{\alpha }\left( \vec{k}\right) }{\partial t} &=&\left(
-\Gamma _{\vec{k},\beta }^{j}\frac{d\vec{k}}{dt}\right) \psi _{\beta }\left( 
\vec{k}\right) , \\
&=&\left( -\left\langle \alpha ,\vec{k}\right\vert \frac{\partial }{\partial 
\vec{k}}\left\vert \beta ,\vec{k}\right\rangle \frac{d\vec{k}}{dt}\right)
\psi _{\beta }\left( \vec{k}\right) \text{.}
\end{eqnarray*}%
In the adiabatic case, this becomes (assuming energy band $\alpha $ is far
remove from the other bands), 
\[
\frac{\partial \psi _{\alpha }\left( \vec{k}\right) }{\partial t}=\left(
-\left\langle \alpha ,\vec{k}\right\vert \frac{\partial }{\partial \vec{k}}%
\left\vert \alpha ,\vec{k}\right\rangle \cdot \frac{d\vec{k}}{dt}\right)
\psi _{\alpha }\left( \vec{k}\right) \text{.} 
\]%
Thus, 
\begin{eqnarray*}
\frac{d\ln \psi _{\alpha }\left( \vec{k}\right) }{dt} &=&\left(
-\left\langle \alpha ,\vec{k}\right\vert \frac{\partial }{\partial \vec{k}}%
\left\vert \alpha ,\vec{k}\right\rangle \cdot \frac{d\vec{k}}{dt}\right) , \\
d\ln \psi _{\alpha }\left( \vec{k}\right) &=&-\left\langle \alpha ,\vec{k}%
\right\vert \frac{\partial }{\partial \vec{k}}\left\vert \alpha ,\vec{k}%
\right\rangle \cdot d\vec{k}, \\
d\phi &=&i\left\langle \alpha ,\vec{k}\right\vert \frac{\partial }{\partial 
\vec{k}}\left\vert \alpha ,\vec{k}\right\rangle \cdot d\vec{k}\text{,}
\end{eqnarray*}%
where $id\phi $ is the change of phase of the wavefunction along a curve in $%
\vec{k}$-space (Brillouin zone). Around a closed curve the total change of
phase must be a multiple of $2\pi $, i.e., $\Delta \phi =2\pi n$ $\left(
n\in 
\mathbb{Z}
\right) $ for the wavefunction to return to its original state. We can write 
\begin{eqnarray*}
&&\frac{i}{\left( 2\pi \right) }\int \int dk_{x}dk_{y}\ f\left( E_{\alpha
}\left( \vec{k}\right) \right) \left[ 
\begin{array}{c}
\left\langle \frac{\partial }{\partial k_{x}}\alpha ,\vec{k}\right\vert 
\frac{\partial }{\partial k_{y}}\left\vert \alpha ,\vec{k}\right\rangle \\ 
-\left\langle \frac{\partial }{\partial k_{y}}\alpha ,\vec{k}\right\vert 
\frac{\partial }{\partial k_{x}}\left\vert \alpha ,\vec{k}\right\rangle%
\end{array}%
\right] , \\
&=&\frac{i}{\left( 2\pi \right) }\int \int dk_{x}dk_{y}\ f\left( E_{\alpha
}\left( \vec{k}\right) \right) \left[ \nabla _{\vec{k}}\times \left\langle
\alpha ,\vec{k}\right\vert \frac{\partial }{\partial \vec{k}}\left\vert
\alpha ,\vec{k}\right\rangle \right] _{plane}.
\end{eqnarray*}%
This result can also be traced to the self-consistent Bohr-Sommerfeld
quantization condition \cite{fego}.

\section{The Kubo current-current correlation formula}

To touch base with a time-dependent perturbation of the Kubo current-current
correlation we recall that in this particular approach, a time varying
electric field is indirectly used. To get to QHE the limiting case of $%
\omega \Longrightarrow 0$ is taken after Fourier transformation of a
convolution integral. In adapting to our approach, this means that the time
integral in the expression of the RHS of Eq. (\ref{pullback}) when
transformed to current-current correlation is a convolution integral before
taking the Fourier transform. We start with the RHS of Eq. (\ref{pullback}), 
\begin{eqnarray}
&&RHS  \nonumber \\
&=&\frac{e^{2}}{h}\frac{1}{\left( 2\pi \hbar \right) }\int \int \int d%
\mathcal{\vec{K}}_{x}d\mathcal{\vec{K}}_{y}dt  \nonumber \\
&&\times \sum\limits_{\alpha ,\beta }\left[ 
\begin{array}{c}
\left( E_{\beta }\left( \mathcal{\vec{K}},\mathcal{E}\right) -E_{\alpha
}\left( \mathcal{\vec{K}},\mathcal{E}\right) \right) \\ 
\times \left\{ 
\begin{array}{c}
\left\langle \alpha ,\frac{\partial }{\hbar \partial k_{x}}\mathcal{\vec{K}},%
\mathcal{E}\right\vert \left\vert \beta ,\mathcal{\vec{K}},\mathcal{E}%
\right\rangle \left\langle \beta ,\mathcal{\vec{K}},\mathcal{E}\right\vert
\left\vert \alpha ,\frac{\partial }{\hbar \partial k_{y}}\mathcal{\vec{K}},%
\mathcal{E}\right\rangle \\ 
-\left\langle \alpha ,\frac{\partial }{\hbar \partial k_{y}}\mathcal{\vec{K}}%
,\mathcal{E}\right\vert \left\vert \beta ,\mathcal{\vec{K}},\mathcal{E}%
\right\rangle \left\langle \beta ,\mathcal{\vec{K}},\mathcal{E}\right\vert
\left\vert \alpha ,\frac{\partial }{\hbar \partial k_{x}}\mathcal{\vec{K}},%
\mathcal{E}\right\rangle%
\end{array}%
\right\} \\ 
\times \ f\left( E_{\alpha }\right) e^{i\left( \omega _{\alpha \beta
}\right) t}%
\end{array}%
\right]  \nonumber \\
&&  \label{RHSpullback}
\end{eqnarray}%
Denoting $E_{\beta }\left( \mathcal{\vec{K}},\mathcal{E}\right) -E_{\alpha
}\left( \mathcal{\vec{K}},\mathcal{E}\right) =\hbar \omega _{\beta \alpha }$%
, we have,%
\begin{eqnarray*}
&&RHS \\
&=&\frac{e^{2}}{h}\frac{1}{\left( 2\pi \hbar \right) }\int \int \int d%
\mathcal{\vec{K}}_{x}d\mathcal{\vec{K}}_{y}dt \\
&&\times \sum\limits_{\alpha ,\beta }\left[ 
\begin{array}{c}
\hbar \omega _{\beta \alpha } \\ 
\times \left\{ 
\begin{array}{c}
\left\langle \alpha ,\frac{\partial }{\hbar \partial k_{x}}\mathcal{\vec{K}},%
\mathcal{E}\right\vert \left\vert \beta ,\mathcal{\vec{K}},\mathcal{E}%
\right\rangle \left\langle \beta ,\mathcal{\vec{K}},\mathcal{E}\right\vert
\left\vert \alpha ,\frac{\partial }{\hbar \partial k_{y}}\mathcal{\vec{K}},%
\mathcal{E}\right\rangle \\ 
-\left\langle \alpha ,\frac{\partial }{\hbar \partial k_{y}}\mathcal{\vec{K}}%
,\mathcal{E}\right\vert \left\vert \beta ,\mathcal{\vec{K}},\mathcal{E}%
\right\rangle \left\langle \beta ,\mathcal{\vec{K}},\mathcal{E}\right\vert
\left\vert \alpha ,\frac{\partial }{\hbar \partial k_{x}}\mathcal{\vec{K}},%
\mathcal{E}\right\rangle%
\end{array}%
\right\} \\ 
\times f\left( E_{\alpha }\right) e^{i\left( \omega _{\alpha \beta }\right)
t}%
\end{array}%
,\right]
\end{eqnarray*}%
\begin{eqnarray*}
&&RHS \\
&=&\frac{e^{2}}{h}\frac{\hbar }{\left( 2\pi \hbar \right) \hbar ^{2}}\int
\int \int d\mathcal{\vec{K}}_{x}d\mathcal{\vec{K}}_{y}dt \\
&&\times \sum\limits_{\alpha ,\beta }\left[ 
\begin{array}{c}
\omega _{\beta \alpha } \\ 
\times \left\{ 
\begin{array}{c}
\left\langle \alpha ,\frac{\partial }{\partial k_{x}}\mathcal{\vec{K}},%
\mathcal{E}\right\vert \left\vert \beta ,\mathcal{\vec{K}},\mathcal{E}%
\right\rangle \left\langle \beta ,\mathcal{\vec{K}},\mathcal{E}\right\vert
\left\vert \alpha ,\frac{\partial }{\partial k_{y}}\mathcal{\vec{K}},%
\mathcal{E}\right\rangle \\ 
-\left\langle \alpha ,\frac{\partial }{\partial k_{y}}\mathcal{\vec{K}},%
\mathcal{E}\right\vert \left\vert \beta ,\mathcal{\vec{K}},\mathcal{E}%
\right\rangle \left\langle \beta ,\mathcal{\vec{K}},\mathcal{E}\right\vert
\left\vert \alpha ,\frac{\partial }{\partial k_{x}}\mathcal{\vec{K}},%
\mathcal{E}\right\rangle%
\end{array}%
\right\} \\ 
\times f\left( E_{\alpha }\right) e^{i\left( \omega _{\alpha \beta }\right)
t}%
\end{array}%
\right] .
\end{eqnarray*}%
We make use of the general relations 
\[
\left\langle \alpha ,\vec{p}\right\vert \vec{v}\left\vert \beta ,\vec{p}%
\right\rangle \equiv \omega _{\beta \alpha }\left\langle \alpha ,\nabla _{%
\vec{k}}\vec{p}\right\vert \left\vert \beta ,\vec{p}\right\rangle 
\]%
\begin{equation}
\left\langle \alpha ,\nabla _{\vec{k}}\vec{p}\right\vert \left\vert \beta ,%
\vec{p}\right\rangle =\frac{\left\langle \alpha ,\vec{p}\right\vert \vec{v}%
_{g}\left\vert \beta ,\vec{p}\right\rangle }{\omega _{\beta \alpha }}
\label{transv1}
\end{equation}%
Similarly, we have 
\[
\left\langle \beta ,\vec{p}\right\vert \vec{v}\left\vert \alpha ,\vec{p}%
\right\rangle \equiv \omega _{\beta \alpha }\left\langle \beta ,\vec{p}%
\right\vert \nabla _{\vec{k}}\left\vert \alpha ,\vec{p}\right\rangle , 
\]%
\begin{equation}
\left\langle \beta ,\vec{p}\right\vert \nabla _{\vec{k}}\left\vert \alpha ,%
\vec{p}\right\rangle =\frac{\left\langle \beta ,\vec{p}\right\vert \vec{v}%
_{g}\left\vert \alpha ,\vec{p}\right\rangle }{\omega _{\beta \alpha }}\text{.%
}  \label{transv2}
\end{equation}%
Substituting in Eq. (\ref{RHSpullback}), we then have the convolution
integral with respect to time, 
\begin{eqnarray*}
&&RHS \\
&=&\frac{e^{2}}{h}\frac{1}{\left( 2\pi \hbar \right) \hbar }\int \int \int d%
\mathcal{\vec{K}}_{x}d\mathcal{\vec{K}}_{y}dt^{\prime } \\
&&\times \sum\limits_{\alpha ,\beta }\left[ 
\begin{array}{c}
\left\{ 
\begin{array}{c}
\frac{1}{\omega _{\beta \alpha }}\left\langle \alpha ,\mathcal{\vec{K}},%
\mathcal{E}\right\vert v_{g,x}\left\vert \beta ,\mathcal{\vec{K}},\mathcal{E}%
\right\rangle \left\langle \beta ,\mathcal{\vec{K}},\mathcal{E}\right\vert
v_{g,y}\left( t-t^{\prime }\right) \left\vert \alpha ,\mathcal{\vec{K}},%
\mathcal{E}\right\rangle \\ 
-\left\langle \alpha ,\frac{\partial }{\partial k_{y}}\mathcal{\vec{K}},%
\mathcal{E}\right\vert v_{g,y}\left( t-t^{\prime }\right) \left\vert \beta ,%
\mathcal{\vec{K}},\mathcal{E}\right\rangle \left\langle \beta ,\mathcal{\vec{%
K}},\mathcal{E}\right\vert v_{g,x}\left\vert \alpha ,\mathcal{\vec{K}},%
\mathcal{E}\right\rangle \frac{1}{\omega _{\beta \alpha }}%
\end{array}%
\right\} \\ 
\times \ e^{i\left( \omega _{\alpha \beta }\right) t^{\prime }}f\left(
E_{\alpha }\right)%
\end{array}%
\right] \text{.}
\end{eqnarray*}%
Consider the following Fourier transformation, 
\[
\frac{1}{\sqrt{2\pi }}\int_{-\infty }^{\infty }e^{i\omega t}F\left( t\right)
dt=\frac{1}{\sqrt{2\pi }}\int_{-\infty }^{\infty }e^{i\omega
t}dt\int_{-\infty }^{0}f\left( t-t^{\prime }\right) g\left( t^{\prime
}\right) dt^{\prime }\text{,} 
\]%
Making the substitution,%
\begin{eqnarray*}
t-t^{\prime } &=&\alpha \Longrightarrow dt=d\alpha \text{,} \\
t &=&\left( t^{\prime }+\alpha \right) \text{,}
\end{eqnarray*}%
we have,%
\[
\frac{1}{\sqrt{2\pi }}\int_{-\infty }^{\infty }e^{i\omega t}F\left( t\right)
dt=\frac{1}{\sqrt{2\pi }}\int_{-\infty }^{\infty }e^{i\omega t^{\prime
}}g\left( t^{\prime }\right) dt^{\prime }\int_{-\infty }^{0}e^{i\omega
\alpha }f\left( \alpha \right) d\alpha \text{.} 
\]%
We can transform the range of integration as follows, 
\begin{eqnarray*}
\int_{-\infty }^{0}e^{i\omega \alpha }f\left( \alpha \right) d\alpha
&=&\int_{\infty }^{0}e^{-i\omega \alpha }f\left( -\alpha \right) \left(
-d\alpha \right) =\int_{0}^{\infty }e^{-i\omega \alpha }f\left( -\alpha
\right) d\alpha , \\
&=&\int_{0}^{\infty }e^{-i\omega \alpha }f^{\dagger }\left( \alpha \right)
d\alpha =\int_{0}^{\infty }e^{-i\omega \alpha }f\left( \alpha \right)
d\alpha \text{,}
\end{eqnarray*}%
since $f\left( \alpha \right) =j\left( \alpha \right) $ is the observable
current density and hence self-adjoint. We can apply this result in what
follows. Defining the current density as%
\[
j_{x}=\frac{ev_{g,x}}{a^{2}}, 
\]
we obtain, after Fourier transforming the convolution integral as 
\begin{eqnarray}
&&RHS  \nonumber \\
&=&\frac{a^{2}}{\hbar \omega }\int \int \left( \frac{a}{\left( 2\pi \hbar
\right) }\right) ^{2}d\mathcal{\vec{K}}_{x}d\mathcal{\vec{K}}%
_{y}\int_{0}^{\infty }dt  \nonumber \\
&&\times \sum\limits_{\alpha ,\beta }\left[ 
\begin{array}{c}
\left\{ 
\begin{array}{c}
\left\langle \alpha ,\mathcal{\vec{K}},\mathcal{E}\right\vert \frac{%
ev_{_{g,}x}}{a^{2}}\left\vert \beta ,\mathcal{\vec{K}},\mathcal{E}%
\right\rangle \left\langle \beta ,\mathcal{\vec{K}},\mathcal{E}\right\vert 
\frac{ev_{g,y}\left( t\right) }{a^{2}}\left\vert \alpha ,\mathcal{\vec{K}},%
\mathcal{E}\right\rangle \\ 
-\left\langle \alpha ,\mathcal{\vec{K}},\mathcal{E}\right\vert \frac{%
ev_{g,y}\left( t\right) }{a^{2}}\left\vert \beta ,\mathcal{\vec{K}},\mathcal{%
E}\right\rangle \left\langle \beta ,\mathcal{\vec{K}},\mathcal{E}\right\vert 
\frac{ev_{g,x}}{a^{2}}\left\vert \alpha ,\mathcal{\vec{K}},\mathcal{E}%
\right\rangle%
\end{array}%
\right\} \\ 
\times \ e^{-i\left( \omega -i\eta \right) t}f\left( E_{\alpha }\right)%
\end{array}%
\right] ,  \nonumber \\
&=&\frac{a^{2}}{\hbar \omega }\int_{0}^{\infty }dt\int \int \left( \frac{a}{%
\left( 2\pi \hbar \right) }\right) ^{2}d\mathcal{\vec{K}}_{x}d\mathcal{\vec{K%
}}_{y}  \nonumber \\
&&\times \sum\limits_{\alpha ,\beta }\left[ 
\begin{array}{c}
\left\{ 
\begin{array}{c}
\left\langle \alpha ,\mathcal{\vec{K}},\mathcal{E}\right\vert j_{x}\left(
0\right) \left\vert \beta ,\mathcal{\vec{K}},\mathcal{E}\right\rangle
\left\langle \beta ,\mathcal{\vec{K}},\mathcal{E}\right\vert j_{y}\left(
t\right) \left\vert \alpha ,\mathcal{\vec{K}},\mathcal{E}\right\rangle \\ 
-\left\langle \alpha ,\mathcal{\vec{K}},\mathcal{E}\right\vert j_{y}\left(
t\right) \left\vert \beta ,\mathcal{\vec{K}},\mathcal{E}\right\rangle
\left\langle \beta ,\mathcal{\vec{K}},\mathcal{E}\right\vert j_{x}\left(
0\right) \left\vert \alpha ,\mathcal{\vec{K}},\mathcal{E}\right\rangle%
\end{array}%
\right\} \\ 
\times \ e^{-i\left( \omega -i\eta \right) t}f\left( E_{\alpha }\right)%
\end{array}%
\right] ,  \nonumber \\
&=&\frac{a^{2}}{\hbar \omega }\int_{0}^{\infty }dt\ Tr\rho _{0}\left\{ \left[
j_{x}\left( 0\right) ,j_{y}\left( t\right) \right] \right\} e^{-i\left(
\omega -i\eta \right) t}\text{,}  \label{kuboformula}
\end{eqnarray}%
where $\eta $ is just a regularization exponent at $\infty $. Therefore the
Kubo formula for the conductivity is given by 
\begin{equation}
\sigma _{yx}\left( t\right) =\frac{a^{2}}{\hbar \omega }\int_{0}^{\infty
}dt\ Tr\rho _{0}\left\{ \left[ j_{x}\left( 0\right) ,j_{y}\left( t\right) %
\right] \right\} e^{-i\left( \omega -i\eta \right) t}\text{.}
\label{kformula2}
\end{equation}%
This is the Kubo current-current correlation formula for the Hall
conductivity.

\bigskip


\begin{thebibliography}{99}
\bibitem{tknn} D. J. Thouless, M. Kohmoto, M. P. Nightingale, and M. den
Nijs, \textit{Quantized Hall Conductance in a Two-Dimensional Periodic
Potential}, Phys. Rev. Lett. \textbf{49}, (6), 405 (1982).

\bibitem{streda} P. Streda, Q\textit{uantised Hall effect in a
two-dimensional periodic potential}, J. Phys. C: Solid State Phys. \textbf{15%
} L1299 (1982).

\bibitem{laughlin} R. B. Laughlin, \textit{Quantized Hall conductivity in
two dimension}, Phys. Rev. B \textbf{23}, 5632 (1981)

\bibitem{halperin} B. I. Halperin, \textit{Quantized Hall conductance,
current-carrying edge states, and the existence of extended states in a
two-dimensional disordered potential}, Phys. Rev. B \textbf{25}, 2185 (1982).

\bibitem{kdp} K. von Klitzing, G. Dorda, and M. Pepper, \textit{A New Method
for High-Accuracy Determination of the Fine--Structure Constant Based on
Quantized Hall Resistance}, Phys. Rev. Lett, \textbf{45}, 494 (1980).

\bibitem{zener} F.A. Buot, "Zener Effect", in Encyclopedia of Electrical and
Electronics Engineering, Ed. John Webster, Vol. 23, pp. 669-688 (John Wiley,
NY 1999). Wiley Online Library 2000 John Wiley \& Sons, Inc.

\bibitem{wannier} G. H. Wannier, "\textit{Dynamics of Band Electrons in
Electric and Magnetic Fields"}, Rev. Mod. Phys. \textbf{34}, 645 (1962).

\bibitem{buotbook} Felix A. Buot, "Nonequilbrium Quantum Transport Physics
in Nanosystems" (World Scientific, 2009) and references therein.

\bibitem{trHn} F. A. Buot, \textit{Method for Calculating }$TrH^{n}$\textit{%
\ in Solid-State Theory}, Phys. Rev. \textbf{B10}, 3700 (1974).

\bibitem{bj} F. A. Buot and K. L. Jensen, "\textit{Lattice Weyl-Wigner
Formulation of Exact Many-Body Quantum Transport Theory and Applications to
Novel Quantum-Based Devices}", Phys. Rev. B\textbf{42}, 9429-9456 (1990).

\bibitem{fego} F.A. Buot, A.R. Elnar, G. Maglasang, and R.E.S. Otadoy, 
\textit{On quantum Hall effect, Kosterlitz-Thouless phase transition, Dirac
magnetic monopole, and Bohr-Sommerfeld quantization}, J. Phys. Commun. 
\textbf{5} 025007 (2021).

\bibitem{Schade} N. B. Schade, D. I. Schuster, and S. R. Nagel, \textit{A
nonlinear, geometric Hall effect without magnetic field}, PNAS December 3,
2019 116 (49) 24475-24479; first published November 18, 2019.

\bibitem{zubkov} M.A. Zubkov and X. Wu, \textit{Topological invariant in
terms of the Green functions for quantum Hall effect in the presence of
varying magnetic field}, Annals of Phys., 168170 (2020). arXiv:1901.06661
(2019).

\bibitem{ikeda} K. Ikeda, \textit{Quantum Hall effect and Langlands program}%
, arXiv:1708.00419v2.

\bibitem{shitade} A. Shitade, \textit{Anomalous thermal Hall effect in
disordered Weyl ferromagnet}, arXiv:1610.00390 (1917).

\bibitem{physicab} F. A. Buot, K. B. Rivero, R. E. S. Otadoy, \textit{%
Generalized nonequilibrium quantum transport of spin and pseudospins:
Entanglements and topological phases}, Physica B: Condensed Matter \textbf{%
559} 42--61 (2019)

\bibitem{fab} F. A. Buot, \textit{Discrete Phase Space and Quantum
Superfield Theory in Nanosystem Quantum Transport}, J. Comput. Theor.
Nanosci. \textbf{4}, 1037-1082 (2007).

\bibitem{fab2} F. A. Buot, \textit{Discrete Phase-Space Model for Quantum
Mechanics}, in M. Kafatos, Ed., Bell's Theorem, Quantum Theory, and
Conceptions of the Universe (Kluwer, NY, 1989, Fundamental Physics Series),
pp. 159-162.

\bibitem{fab3} F. A. Buot, \textit{Direct Construction of Path Integrals in
the Lattice-Space Multiband Dynamics of Electrons in a Solid}, Phys. Rev. 
\textbf{A33}, 2544- 2562(1986).

\bibitem{fab4} F. A. Buot, \textit{General Theory of Quantum Distribution
Function Transport Equations: Superfuid Systems and Ultrafast Dynamics of
Optically Excited Semiconductors}, La Rivista del Nuovo Cimento \textbf{20},
No.9, 1-75 (1997).

\bibitem{fab5} Felix A. Buot,\textquotedblright\ Foundation of computational
nanoelectronics\textquotedblright , in Handbook of Theoretical and
Computational Nanotechnology, American Scientific Publishers (2006), Vol. 1.
pp. 221-310.

\bibitem{fab6} Felix A. Buot, \textit{Operator Space and Discrete Phase
Space Methods in Quantum Transport and Quantum Computing}, J. Comp. Theor.
Nanoscience \textbf{6},1864-1926 (2009).

\bibitem{gibbons} K.S. Gibbons, M.J. Hoffman, W.K. Wootters, \textit{%
Discrete phase space based on finite fields}, Phys. Rev. \textbf{A70},
062101 (2004).

\bibitem{kasper} P. Kasperkovitz and M . Peev, \textit{Wigner-Weyl
Formalisms for Toroidal Geometries}, Annals of Physics \textbf{230}, 21-51
(1994).

\bibitem{liga} M. Ligabo, \textit{Torus as phase space: Weyl quantization,
dequantization, and Wigner formalism}, J. Math. Phys. \textbf{57}, 082110
(2016).

\bibitem{fialkovs} I.V.Fialkovsky and M.A.Zubkova, \textit{Precise
Wigner-Weyl calculus for lattice models}, Nuclear Phys. B \textbf{954},
114999 (2020).

\bibitem{comments} Felix A. Buot, \textit{Comments on the Weyl-Wigner
calculus for lattice models}, {\small http://arxiv.org/abs/2103.10351}

\bibitem{bok} Felix A. Buot, Roland E. S. Otadoy, and Karla B. Rivero, 
\textit{Magnetic susceptibility of Dirac fermions, BiSb alloys, interacting
Bloch fermions, dilute nonmagnetic alloys, and Kondo alloys}, Physica B 
\textbf{503}, 69-97 (2017).

\bibitem{fabis} F.A. Buot, Formalism of distribution-function method in
impurity screening, Phys. Rev B \textbf{13}, 977-989 (1976).

\bibitem{jb} K.L. Jensen and F.A. Buot, \textit{Numerical simulation of
intrinsic bistability and high-frequency current oscillations in resonant
tunneling structures}, Phys. Rev. Lett. \textbf{66}, 1078 (1991).

\bibitem{rossi} F. Rossi, A.Di Carlo, and P. Lugli, \textit{Microscopic
theory of quantum transport phenomena in mesoscopic systems: A Monte Carlo
approach}, Phys. Rev. Lett. \textbf{80}, 3348 (1998).
\end{thebibliography}
\end{document}